\documentclass{article}[12pt]
\pdfoutput=1
\usepackage{amsfonts}       
\usepackage{amsmath}        
\usepackage{amssymb}
\usepackage{epsfig}

\def\be{\begin{equation}}
\def\ee{\end{equation}}
\def\bea{\begin{eqnarray}}
\def\eea{\end{eqnarray}}
\def\ocal{\mathcal{O}}

\def\fcal{\mathcal{F}}
\def\pcal{\mathcal{P}}
\def\jcal{\mathcal{J}}
\def\w{\omega}
\def\pa{\partial}
\def\R{\mathbb{R}}
\def\fw{{\frak{w}}}
\def\fk{{\frak{q}}}
\def\PD{{\mathcal P}}
\def\CD{{\mathcal C}}
\def\TD{{\mathcal T}}
\def\half{{\textstyle{1\over2}}}
\def\ang{\mbox{\AA}}

\begin{document}

\begin{flushright}
NSF-KITP-07-207 \\
PUPT-2254 \\
arXiv:0801.1693 [hep-th]
\end{flushright}

\begin{center}
\vspace{1cm} { \LARGE {\bf Impure AdS/CFT}}

\vspace{1.1cm}

Sean A. Hartnoll$^\flat$ and Christopher P. Herzog$^\sharp$

\vspace{0.8cm}

{\it $^\flat$ KITP, University of California\\
     Santa Barbara, CA 93106-4030, USA }

\vspace{0.8cm}

{\it $^\sharp$ Department of Physics, Princeton University \\
     Princeton, NJ 08544, USA }

\vspace{0.8cm}

{\tt hartnoll@kitp.ucsb.edu, cpherzog@Princeton.EDU} \\

\vspace{2cm}

\end{center}

\begin{abstract}
\noindent
We study momentum relaxation due to dilute, weak impurities in a
strongly coupled CFT, a truncation of the M2 brane theory. Using
the AdS/CFT correspondence, we compute the relaxation timescale as
a function of the background magnetic field $B$ and charge density
$\rho$. The theory admits two different types of impurities. We
find that for magnetic impurities, momentum relaxation due to the
impurity is suppressed by a background $B$ or $\rho$. For electric
impurities, due to an underlying instability in the theory towards
an ordered phase, the inverse relaxation timescale increases
dramatically near $\sqrt{B^2 + \rho^2/\sigma^2_0} \sim 21 T^2$. We
compute the Nernst response for the impure theory, and comment on
similarities with recent measurements in superconductors.

\end{abstract}

\pagebreak
\setcounter{page}{1}

\section{Introduction}

Second order quantum phase transitions can occur at zero
temperature in condensed matter systems when there is a
non-analytic change in the ground state energy as a function of
some coupling \cite{sachdev}. On either side of the critical
point, at sufficiently low temperatures the system admits a
quasiparticle description. The nature of the quasiparticles
depends on the orders characterising the different phases. For
instance, N\'eel ordered phases have spin wave excitations whereas
phases with Valence Bond Solid order have spinon and `photon'
excitations \cite{sachdev2}. However, as the temperature is
increased, near to the critical coupling, thermal fluctuations
lead to competition between the different orders. In this strongly
coupled regime, in which both quantum and thermal fluctuations
play an important role, neither quasiparticle description is
appropriate. Instead the system is best described by a finite
temperature 2+1 dimensional conformal field theory (CFT)
\cite{sachdev, sachdev2}. Often this theory turns out to be
relativistically invariant (the `speed of light' in these theories
is not $c$ but rather some lower speed characterising the
material).

Quantum critical points are believed to be important in various
systems of experimental interest, including the high $T_{c}$
cuprate superconductors \cite{tranquada, sachdev3}. In contrast,
the theorist's toolkit of tractable 2+1 CFTs has largely been
limited to the $O(N)$ model at large $N$, and related models with
relevant quartic interactions \cite{sachdev, sachdev2}. The
AdS/CFT correspondence \cite{Maldacena:1997re} provides a wealth
of new examples of 2+1 CFTs in which explicit computations are
possible at large $N$. Although this correspondence has been
intensely studied for a decade, potential applications to concrete
condensed matter systems have only recently begun to be explored
\cite{Herzog:2007ij, Hartnoll:2007ai, Hartnoll:2007ih, HartnollHerzog}.
Relevant earlier work on finite temperature physics in $AdS_{4}$
includes \cite{Herzog:2002fn, Herzog:2003ke, Saremi:2006ep, japanese}. Most
immediately these CFTs provide new solvable toy models for
strongly coupled dynamics in 2+1 dimensions. In the future it
might be possible to engineer the relevant supersymmetric CFTs in
a lab. One can thus speculate that in addition to having
interesting physical properties in their own right, these systems
open up the possibility of doing experimental quantum gravity via
the AdS/CFT correspondence.

In many experimental systems, microscopic impurities in the
samples leave a significant imprint on the physics. Most directly,
the impurities are at fixed (random) locations and therefore break
translational invariance. This loss leads to a relaxation of
momentum at late times. Momentum relaxation is characterised by a
timescale $\tau_{\text{imp}}$, which will be the main focus of
this paper. In many contexts, see e.g.\ \cite{Hartnoll:2007ih} for
a discussion, momentum relaxation is necessary for dc ($\w=0$)
transport to be finite; otherwise the `Drude peak' becomes a delta
function as $1/\tau_{\text{imp}} \to 0$. We will exhibit similar
`metallic' behaviour in a strongly coupled CFT with impurities.

Given a theory, the effect of impurities can in principle be
computed microscopically. One would like to know how the impurity
timescale depends on temperature, background magnetic field and
charge density. To our knowledge there is no first principles
computation of $\tau_{\text{imp}}(T,B,\rho)$ available in the
literature. If the impurities are sufficiently dilute and
furthermore weak enough for their interaction with the CFT to be
treated perturbatively, then $\tau_{\text{imp}}$ can be computed
from a two point function in the finite temperature CFT. We will
derive the formula for $\tau_{\text{imp}}$ and use the AdS/CFT
correspondence to compute it for a truncation of the M2 brane
theory.

In sections 2 and 3 we give a general description of the effect of
dilute, weak impurities in a CFT. The main result of these
sections is equation (\ref{eq:tauimp}), which expresses the
impurity relaxation timescale in terms of the retarded Green's
function of a relevant scalar operator in the theory which couples
to the impurities.

In section 4, we show how this Green's function may be evaluated
in a truncation of the M2 brane theory using the AdS/CFT
correspondence. Our truncation is not consistent.  We neglect a
coupling between the scalar and pseudoscalar modes we keep and a
second gauge field. The inconsistent truncation is made to keep
the calculations simple and also because we have no good physical
interpretation of the second gauge field in our condensed matter
context. In another small abuse of the M2 brane theory, we give
our pseudoscalar operator conformal dimension one instead of two.
This gives us a larger collection of relevant operators to study.
Note however that the standard M2 brane theory is obtainable as a
double trace deformation of ours
\cite{Witten:2001ua}.

In the following section 5 we give our (numerical)
results for $\tau_\text{imp}(T,B,\rho)$. We first note that there
are in fact two relevant operators in the theory, dual to linear
combinations of the scalar and pseudoscalar of the bulk, with
considerably different effects. The operators can be distinguished
by their transformation properties under charge conjugation $\CD$,
parity $\PD$, and time reversal $\TD$. In the presence of a
background magnetic field and finite charge density, the first
operator, $\ocal_+$, transforms the same way as the magnetic field
while $\ocal_-$ transforms as the electric field. We are thus
tempted to identify $\ocal_+$ with magnetic impurities and
$\ocal_-$ with electric impurities.

For both operators, we find that the impurity time
$1/\tau_\text{imp}$ depends on the charge density $\rho$ and
magnetic field $B$ only in the combination $\sqrt{B^2 +
\rho^2/\sigma_0^2}$ where $\sigma_0$ is the electrical conductivity at
$\rho=B=0$.  This combined dependence, while remarkable, may
be an accident of the details of the M2 brane theory.

For the $\ocal_+$ impurities, we find that increasing the magnetic
field or charge density suppresses momentum relaxation due to
impurities. At the same time, at nonzero $B$ there is an
independent source of momentum relaxation due to hydrodynamic
cyclotron motion. Increasing $B$ increases the momentum relaxation
due to cyclotron motion.

However, for the other operator, $\ocal_-$, momentum relaxation
becomes faster with increasing $B$ or $\rho$. When
$\sqrt{B^2+\rho^2/\sigma_0^2} $ reaches a critical value of about
$21 T^2$, then $1/\tau_\text{imp} \to \infty$. We will see that
this divergence is symptomatic of an underlying instability of the
theory towards an ordered phase. From a gravitational point of
view, the instability occurs when the scalar field dual to
$\ocal_-$ develops an unstable mode. This instability provides a
clean counterexample to the original version of the Gubser-Mitra
`correlated stability conjecture'
\cite{Gubser:2000mm,Gubser:2000ec}, similar to those discussed in
\cite{FGM}.

In section 6 we use our results to compute the Nernst coefficient
in the CFT. We note that our plots show some qualitative
similarities to the organic superconductors studied in
\cite{ardavan}, perhaps indicating a nearby quantum critical point
in that system. While the organic superconductors effectively have
vanishing charge density, $\rho=0$, because they are not doped, in
the cuprate superconductors the doping $x$ can be thought of as a
nonzero charge density \cite{Hartnoll:2007ih}.

\section{The effect of random disorder}

The presence of impurities breaks the translational and
conformal invariance of the CFT. Locally the impurities will
source relevant operators and potentially drive the
theory away from the fixed point. Thus we can model the impurities by
adding the following coupling to the Hamiltonian
\be\label{eq:deltaH}
\delta H = \int d^2y V(y) \ocal(t, y) \,.
\ee
The operator $\ocal(t,y)$ is the most relevant operator in the
conformal field theory that preserves the global symmetries of the
theory. Charged impurities are also of interest, although we shall
only study neutral impurities here. This term breaks translation
invariance, because the potential $V(y)$ is explicitly space
dependent, and is hard to work with directly. To get around this
one treats the impurity potential statistically. If the impurities
are sufficiently dilute that their effects do not overlap, then
the precise weighting on the space of potentials is not important,
and it is useful to take the Gaussian
\cite{simons}
\be
\langle \cdots \rangle_{\text{imp}} = \int {\mathcal D} V e^{- \int d^2y V(y)^2 / 2 \bar V^2} (\cdots) \,,
\ee
which implies
\be\label{eq:potential}
\langle V(x) \rangle_\text{imp} = 0 \,, \qquad
\langle V(x) V(y) \rangle_\text{imp} = \bar V^2 \delta^{(2)}(x-y) \,.
\ee
We will furthermore make the approximation that scattering off
impurities may be treated perturbatively, so that we can expand in
powers of $V$. The strength of the potential $\bar V$ is a
dimensionful quantity, with mass dimension
\be
\left[ \bar V \right] = 2 - \Delta_\ocal \,.
\ee
So long as $2 -\Delta_\ocal > 0$, $\bar V$ has positive scaling dimension
and the impurities are a relevant perturbation.  This condition that
$\bar V$ has positive scaling dimension is often called the Harris
criterion \cite{Cardybook}. In the case of the M2 brane theory, the most relevant neutral
operators are mass terms for the scalar fields
\cite{Aharony:1998rm, Minwalla:1998rp, Halyo:1998mc}. By `neutral' we mean with respect
to a certain $U(1) \subset SO(8)$, as we describe below. These
operators have conformal dimension $\Delta_\ocal = 1$. We can
anticipate that the relevance of these operators implies that they
will become strongly coupled at low energies. We will be careful
to keep track of the regime in which perturbation theory in $\bar
V$ is valid.

The breaking of translation invariance leads to the late time non
conservation of momentum. The timescale associated with the loss
of momentum is called the impurity relaxation timescale and is
given to leading order in strength of the impurity potential, $V$,
by
\be\label{eq:tau}
\frac{1}{\tau_\text{imp}} = - \frac{1}{\chi_0} \lim_{\w \to 0} \frac{\text{Im} \, G^R_{\fcal \fcal}(\w,0)}{\w} \,.
\ee
This equation follows from considering the `memory function'. We
give a derivation in the following section, along with a
discussion of the precise meaning of $\tau_\text{imp}$ and of
regimes of validity. To fix conventions, we define the retarded
Green's function of any operator $\ocal$ in Fourier space to be
\be
G^R_{\ocal \ocal}(\w,k) = - i \int d^2x  \int_0^\infty dt \,
\langle [\ocal(t,x), \ocal(0,0)] \rangle e^{i \w t - i k \cdot x}
\,.
\ee
The quantities that appear in (\ref{eq:tau}) are firstly
\be\label{eq:chi}
\chi_0 \equiv \lim_{\omega \to 0} G^R_{\pcal \pcal}(\omega,0) = \epsilon + P \,,
\ee
where $\pcal$ is the momentum density in a fixed direction, i.e.
$\pcal = n_i T^{0 i}$ for some unit vector $n \in \R^2$.  The
number $\chi_0$ is the static susceptibility for the momentum
density, $\epsilon$ is the energy density and $P$ the pressure.
The remaining term in (\ref{eq:tau}) is the Green's function for
the operator
\be\label{eq:fcal}
\fcal(t,x)  =  [\pcal(t,x), \delta H] = \int d^2y \, V(y) [\pcal(t,x), \ocal(t,y)] \ .
 \ee
 This expression
simplifies using that the spatial integral of $\pcal(t,x)$ is a momentum and
hence generates translations.  Locally therefore
$[\pcal(t,x), \ocal(t,y)] = i \delta^{(2)}(x-y) \pa \ocal(t,y)$
up to a total derivative with respect to $x$.
We shall ignore this total derivative; one can use translation invariance
to show that this derivative does not contribute
to the Green's function $G_{\fcal \fcal}^R$.  Thus (\ref{eq:fcal}) becomes
 \be
 \fcal(t,x)  =  i \int d^2y V(y) \delta^{(2)}(x-y) \pa \ocal(t,y) = i V(x) \pa \ocal(t,x) \,.
\ee
Here $\pa$ denotes spatial derivative in the same direction as
$\pcal$, i.e. $\pa = n^i \pa_i$.

The Green's function is therefore
\be\label{eq:withV}
G^R_{\fcal \fcal}(\w, 0) = i \bar V^2 \int_0^\infty dt \langle
[\pa \ocal (t,0), \pa \ocal(0,0) ] \rangle e^{i \w t} \,,
\ee
where we used (\ref{eq:potential}). Passing to momentum space, one obtains
\be
G^R_{\fcal \fcal}(\w, 0) = - \frac{\bar V^2}{2} \int
\frac{d^2k }{(2\pi)^2} k^2 G^R_{\ocal \ocal}(\w,k) \,.
\ee
The factor of $1/2$ arises because the integral with $(n \cdot
k)^2$ in the integrand is half the integral with $k^2$, by
isotropy. Thus
\be\label{eq:tauimp}
\fbox{$\displaystyle
\frac{1}{\tau_{\text{imp}}} =  \frac{\bar V^2}{2 \chi_0} \lim_{\w \to
0} \int \frac{d^2k }{(2\pi)^2}\, k^2 \frac{\text{Im} G^R_{\ocal
\ocal}(\w,k)}{\w} \,.$}
\ee
This is the formula we will use to compute the relaxation
timescale $\tau_\text{imp}$. We will obtain the dependence on
charge density $\rho$ and background magnetic field $B$. Due to
conformal invariance and dimensional analysis, one has the scaling
form
\be\label{eq:tscaling}
\frac{1}{\tau_\text{imp}} = \frac{\bar V^2}{T^{3 - 2
\Delta_\ocal}} F\left(\frac{\rho}{T^2}, \frac{B}{T^2} \right) \,.
\ee

\section{The impurity timescale}

\subsection{Impurity in the absence of a magnetic field}
\label{sec:impurities}

In this section we derive the expression (\ref{eq:tau}) for the
impurity timescale. First note that from the definition of the
retarded Green's function, and (twice) using the Heisenberg
equation of motion one obtains
\be
\w^2 G^R_{\pcal \pcal}(\w,k) =  - G^R_{[\pcal,H] [\pcal,H]}(\w,k)
+ G^R_{[\pcal,H] [\pcal,H]}(0,k)  \,.
\ee
This equation is true up to contact terms for any operator. In the
absence of the impurity potential, $\pcal$ is the conserved
momentum density as well as the energy current. If we split the
total Hamiltonian as $H = H_0 + \delta H$, with $\delta H$ given
by (\ref{eq:deltaH}) above, then $H_0$ is the Hamiltonian of a
translationally invariant theory while $\delta H$ explicitly
breaks this symmetry. It follows that $[\int d^2x \, \pcal(x),H_0]
= 0$. Thus from the definition of the retarded Green's function,
at zero spatial momentum we have that $G^R_{[\pcal,H_0]
\ocal}(\w,0) = 0$, for any operator $\ocal$.
It follows that
\be\label{eq:secderiv}
\w^2 G^R_{\pcal \pcal}(\w,0) = - \left(G^R_{\fcal \fcal}(\w,0)
- G^R_{\fcal \fcal}(0,0)\right) \,.
\ee
Evaluated in a background with a nonzero $\bar V$, this result for
$G_{\pcal \pcal}^R$ is exact.  For us the usefulness of this
expression lies in the fact that a $\bar V^2$ trivially factors
out of the right hand side, see equation (\ref{eq:withV}),
allowing us to evaluate the remainder in a background with
vanishing $\bar V$, thus yielding a result for $G_{\pcal \pcal}^R$
that is accurate to leading order in $\bar V^2$.

Given this formula (\ref{eq:secderiv}), $\lim_{\bar V \to 0}
G^R_{\pcal\pcal}(\omega,0) =0$ as required by the restoration of
translational invariance in this limit. However, we have no
guarantee that at fixed nonzero $\bar V$ the $\lim_{\omega \to 0}
G^R_{\pcal\pcal}(\omega,0)/\omega$, which is proportional to a dc
thermal conductivity, is finite. Yet physically, we expect the
system with impurities to be well behaved in response to low
frequency perturbations.  These considerations suggest that the
response of the system is governed by a modified Green's function
where a contact term has been subtracted:
\be\label{eq:gsubs}
i \theta(\omega,0) T \omega \equiv G^R_{\pcal \pcal}(\omega, 0) -
\chi_0.
\ee
The quantity $\theta(\omega,0)$ might be called a momentum
conductivity; in appendix \ref{sec:transport} we show that indeed $\langle \pcal
\rangle = - \theta(\w,0) \nabla T$. 
 By $\theta$ we mean a
diagonal component of the conductivity matrix; there is no
off-diagonal conduction in the absence of a magnetic field.

We can now read off $\chi_0$ in (\ref{eq:gsubs}). In a
translationally invariant system in the absence of a magnetic
field, the momentum conductivity $\theta(\omega,0) = i(\epsilon+P)
/ \omega T$ where $\epsilon$ is the energy density and $P$ the
pressure. This result can be seen from hydrodynamics
\cite{Hartnoll:2007ih} or from the Ward identity
arguments in \cite{HartnollHerzog}. The fact that the imaginary
part of the conductivity behaves like $1/\w$ implies that the real
part will contain a $\delta(\w)$, because with some suitable
$i\epsilon$ prescription $1/\w = P(1/\w) - i \pi \delta(\w)$. This
delta function is expected for the energy flow in a
translationally invariant system, which has no way to dissipate
momentum. From the fact that translation invariance implies
$G^R_{\pcal\pcal}(\omega,0)=0$, it follows that
\be
\chi_0 = \epsilon+P  \,.
\ee
That $\chi_0 = \lim_{\w \to 0} G^R_{\pcal\pcal}(\omega,0)$ remains
finite as $\bar V \to 0$ is consistent with the fact that the $\w
\to 0$ and $\bar V \to 0$ limits of $G^R_{\pcal\pcal}(\omega,0)$
do not commute\footnote{%
 In the absence of an external
 electric field, $\theta$ is related to the thermal conductivity
 $\bar \kappa$ and thermoelectric coefficient $\hat \alpha$
 defined, for example, in ref.~\cite{HartnollHerzog} via $\theta =
 \bar \kappa + \hat \alpha \mu$.  Thus at vanishing chemical potential $\mu$,
 the thermal conductivity is $\theta$. 
 With no chemical potential, then $\chi_0 = s T$,
 with $s$ the entropy density, leading to $\bar \kappa = is /\w$.
}.

\subsubsection*{The memory function method}

Now we extract the timescale from $G^R_{\pcal \pcal}(\w,0)$. The
memory function $M(\w)$ is defined by
\be\label{eq:memory}
M(\omega) \equiv \frac{\omega G^R_{\pcal \pcal}(\w,0)}{\chi_0 - G^R_{\pcal \pcal}(\w,0)} \ ,
\ee
which can be formally inverted to yield
\be\label{eq:greens}
G^R_{\pcal \pcal}(\w,0) = \frac{\chi_0 M(\w)}{\w + M(\w)} \,.
\ee
The reason for introducing this function is to reinterpret the
small $\omega$ limit of $M(\omega)$ as an inverse scattering time
$i /\tau_\text{imp}$. We comment on this interpretation below. At
finite $\omega$, by translational invariance we expect $G^R_{\pcal
\pcal}(\w,0)$ to vanish in the absence of impurities.  We also
expect that the inverse scattering time goes to zero with no
impurities. Thus, consistent with eq.~(\ref{eq:memory}), we take
both $M(\omega)$ and $G^R_{\pcal
\pcal}(\w,0)$ to scale as $\bar V^2$. These statements do not hold
when there is a nonvanishing background magnetic field. For the
moment we are setting $B=0$. As we noted, $\chi_0 = G^R_{\pcal
\pcal}(0,0)$ is independent of $\bar V$ to leading order. Thus
from eq.~(\ref{eq:memory}), and using equation
(\ref{eq:secderiv}), we find the approximate expression for
$M(\w)$:
\be\label{eq:mw}
M(\w) \approx \frac{\omega G^R_{\pcal \pcal}(\w,0)}{\chi_0} =
\frac{- \left(G^R_{\fcal \fcal}(\w,0)
- G^R_{\fcal \fcal}(0,0)\right) / \omega}{\chi_0} \ .
\ee
The approximation is up to terms of ${\mathcal O}(\bar V^3)$. We
would like to take the small $\w$ limit of this expression.
Because $\bar V$ is dimensionful, one should expect that in order
to obtain the true $\w \to 0$ limit it will be necessary to resum
higher order contributions in $\bar V/\w^{2-\Delta_\ocal}$.
Indeed, we have already noted that the $\w
\to 0$ and $\bar V \to 0$ limits do not commute. Using
(\ref{eq:mw}) directly to compute the relaxation timescale, is
often called the `memory function method', see for instance
refs.~\cite{gotze, giamarchi}. It was shown in ref.~\cite{argyres}
that this method generally does not give the correct answer for
small frequencies $\omega^{2-\Delta_\ocal} < \bar V$.

We can obtain reliable results in the case when the temperature is
large compared to the strength of the scattering potential: $\bar
V^{1/(2 - \Delta_\ocal)} \ll T$. Given this separation of scales,
at low frequencies $\w \ll T$,  the constraint
$\omega^{2-\Delta_\ocal} > \bar V$, for the validity of
perturbation theory, should be replaced by the weaker constraint
$T^{2-\Delta_\ocal} > \bar V$. In the language of the
renormalization group, $\bar V$ is a relevant operator and we must
choose a scale at which to evaluate it. At nonzero temperature and
at frequencies $\omega < T$, we expect the temperature to act as a
cut off in the renormalization flow for $\bar V$. In higher order
corrections to the memory function $M$, the ratio $\bar V / T^{2 -
\Delta_\ocal}$ will appear in place of $\bar V /
\omega^{2 - \Delta_\ocal}$. Thus at high temperatures, we are justified in
treating the impurities perturbatively, including at very low
frequencies.

Evaluated perturbatively, the expression for the memory function
(\ref{eq:mw}) has an overall factor of $\bar V^2$, and no other
dependence on $\bar V$. Scaling therefore implies that the
frequency dependence must be some function of $\w/T$. Thus the
hydrodynamic limit $\w \ll T$ is equivalent to taking $\w
\to 0$ in (\ref{eq:mw}). We can identify
\begin{eqnarray}
\frac{i}{\tau_\text{imp}} &=& \frac{1}{\chi_0} \lim_{\omega \to 0}
\frac{- \left(G^R_{\fcal \fcal}(\w,0)
- G^R_{\fcal \fcal}(0,0)\right)}{\omega} \\
&=& -\frac{1}{\chi_0} \left. \frac{d G^R_{\fcal \fcal}(\omega,0)}{d \w} \right|_{\omega=0}
\nonumber
\ .
\end{eqnarray}
Using the fact that $\mbox{Re} \, G^{R}_{\fcal \fcal}$ is an even
function while $\mbox{Im} \, G^{R}_{\fcal \fcal}$ is an odd
function of $\omega$, we find our main result of the section that
\be\label{eq:timescale}
\fbox{$\displaystyle
\frac{1}{\tau_\text{imp}} = -\frac{1}{\chi_0} \lim_{\w \to 0}
\frac{\mbox{Im} \, G^{R}_{\fcal \fcal}(\omega,0)}{\w} \ .$}
\ee

Let us clarify the physical meaning of $\tau_\text{imp}$.
Strictly, the momentum relaxation timescale is given by the
imaginary part of the pole in $G^{R}_{\pcal \pcal}(\omega,0)$ that
is closest to the real axis. The identification of the zero
frequency limit as the inverse relaxation timescale
\be
\frac{i}{\tau_\text{imp}} = \lim_{\omega \to 0} M(\w) \,,
\ee
is valid so long as the putative pole at $-i/\tau_{\text{imp}}$ is
sufficiently close to the real axis, in the sense that
\be\label{eq:condition}
\left| M\left(\frac{-i}{\tau_\text{imp}}\right) -
\frac{i}{\tau_\text{imp}} \right|
\ll \frac{1}{\tau_\text{imp}} \, .
\ee
It is easy to check, for instance, that this will be true if
$G^{R}_{\pcal \pcal}(\omega,0)$ has several poles, but one is much
closer to the real axis than the others. Given that
$1/\tau_\text{imp} \sim
\bar V^2/T^{3-2\Delta_\ocal} \ll T$ is parametrically small compared
to the temperature scale, we expect that (\ref{eq:condition})
holds. However, if the coefficient of $\bar V^2$ in
(\ref{eq:tauimp}) becomes sufficiently large, then that expression
is no longer reliable as a relaxation timescale.

We argued above that at high temperatures, potentially dangerous
$\bar V /\omega^{2 - \Delta_\ocal}$ corrections would be replaced
by corrections in $\bar V / T^{2 - \Delta_\ocal}$. While $\bar V /
\omega^{2 -
\Delta_\ocal}$ corrections would tend to add more poles close to
the origin of the complex $\omega$ plane, $\bar V / T^{2 -
\Delta_\ocal}$ corrections are suppressed.

\subsection{Impurity plus an external magnetic field}
\label{sec:mag}

There is great formal similarity between adding to the Lagrangian
a random potential coupled to a scalar operator $\int d^3x \, V(x)
{\mathcal O}(x)$ and adding an external $U(1)$ gauge field coupled
to a global current $-\int d^3x \, A_\mu(x) J^\mu(x)$. This second
case was considered in detail in ref.~\cite{HartnollHerzog}.
A constant magnetic field has $F {=} dA {=} B \, dx \wedge dy$.
Any potential $A$ for this field strength will formally break
translation invariance, which is one way to see why
$G^R_{\pcal\pcal}(\omega,0)$ can be nonzero in the absence of
impurities but in the presence of a background magnetic field.

From the Ward identity results in ref.~\cite{HartnollHerzog}, with
$\bar V = 0$ one has
\be\label{wardjj}
\w G^R_{\pcal\pcal}(\omega,0) = B^2 \left(G^R_{\jcal\jcal}(\omega,0)
- G^R_{\jcal\jcal}(0,0) \right) /\w\,,
\ee
where $\jcal = J^i n_i$ is the electric current.
The $\w \to 0$ limit of the current-current
Green's function needs to be taken with care, as the $B \to 0$ and
$\w \to 0$ limits do not commute.

At $\bar V=0$, $G^R_{\jcal \jcal}(0,0)$ vanishes for real $n_i$,
but there is still a Hall conductivity \cite{Hartnoll:2007ai, HartnollHerzog}, 
\be
\lim_{\w \to 0} \frac{G^R_{J^x J^y}(\w,0)}{\w} =
 i \rho / B \ .
\ee
To see the Hall effect within our formalism, it is convenient to
let the $n_i$ be complex and think of $G^R$ as an inner product.
This method of complexifying the conductivities was found to be
useful in \cite{HartnollHerzog}. In this paper we will do so for
this section and appendix \ref{sec:hydrotau} only. Taking $n_i = (1,-i)/\sqrt{2}$, we find
\be
2 G^R_{\jcal \jcal} = G^R_{J^x J^x} - i G^R_{J^x J^y} + i G^R_{J^y J^x} + G^R_{J^y J^y} \ .
\ee
By rotational invariance, this expression simplifies to
\be
G^R_{\jcal \jcal} = G^R_{J^x J^x} - i G^R_{J^x J^y}  \ .
\ee

Combining eqs.~(\ref{wardjj})
and (\ref{eq:secderiv}) then to include both impurities and a
magnetic field leads to
\be\label{eq:total}
\w^2 G^R_{\pcal\pcal}(\omega,0) =  - \left(G^R_{\fcal \fcal}(\w,0)
- G^R_{\fcal \fcal}(0,0)\right) + B^2
\left(G^R_{\jcal\jcal}(\omega,0) -  \omega \rho /B
\right) \,.
\ee
Like eq.~(\ref{eq:secderiv}), we believe this expression is exact
when the right hand side is evaluated in a background with
nontrivial $B$ and $\bar V$ and 
$G^R_{\jcal \ocal}=0$. We have been loose in our derivation,
but the result can be made rigorous through a Ward identity
argument. 
The potential cross term proportional to $G^R_{\jcal
\ocal}$ will vanish anyway to leading order in $\bar V$
after averaging over $V$.
But in fact, we will see shortly that the Green's function
$G^R_{\jcal \ocal} = 0$ vanishes for the M2 brane theory at
leading order in $1/N$ before averaging.

We call poles hydrodynamic if they are close to the origin of the complex
$\omega$ plane compared to the scales $T$ and $\rho$.
In the previous section, in the absence of a $B$ field,
we argued that for small $\bar V$ there was a hydrodynamic pole in
$G_{\pcal \pcal}^R$ at ${-}i/\tau_{\rm{imp}}$.
In ref.~\cite{HartnollHerzog}, we saw that in the presence of a small magnetic field but in
the absence of impurities, there are a pair of hydrodynamic poles at
 \be\label{eq:cyclo}
\pm \w_c - i \gamma = \frac{B \left(\pm \rho - i \sigma B\right)}{\epsilon + P} \,,
\ee
both in $G_{\pcal \pcal}^R$ and $G_{\jcal \jcal}^R$. These poles
correspond to damped relativistic cyclotron motion\footnote{%
Again, our time dependence is $e^{-i \omega t}$.  Thus $\gamma>0$
and $\tau_\text{imp}>0$ lead to exponential damping at long times.
}. 
Here $\sigma$ is the electrical conductivity of the CFT.

Physically, if we have a small but nonzero $\bar V$ or $B$, we
expect the $\omega \to 0$ response of the system to be well
behaved. The poles in the Green's functions should interpolate
continuously between the cases in which either $\bar V$ or $B$
vanish. Thus, contemplating (\ref{eq:total}), it is natural to
expect that the hydrodynamic poles in the presence of both $\bar
V$ and $B$ are
\be\label{eq:memorytot}
- i \tau_\text{imp}^{-1} -i \gamma \pm \w_c
\,.
\ee
This combination of $\tau_\text{imp}$ and the cyclotron pole was
also found in the hydrodynamic analysis of
\cite{Hartnoll:2007ih}. In eq.~(\ref{eq:memorytot}) we see
that there are two sources for momentum relaxation; one is
impurities and the other can na\"ively be thought of as due to
collisions of positively and negatively charged excitations of the
theory undergoing cyclotron motion in opposite directions.
In appendix \ref{sec:hydrotau}, we demonstrate how the
memory function method can be used to extract the correct poles
from the hydrodynamic
expressions for $G_{\pcal \pcal}$ and $G_{\jcal \jcal}$, and we also
show that these hydrodynamic expressions satisfy the
relation (\ref{eq:total}).

The impurity relaxation timescale $\tau_\text{imp}$ appearing in
eq.~(\ref{eq:memorytot}) is given by the same formula as before,
eq.~(\ref{eq:timescale}), but can now have a $B$ dependence,
provided $B$ is kept small so that our expression for the
cyclotron pole is reliable.  In more detail, our hydrodynamic
expression for $\gamma$ is correct to order $B^2$ but we expect it
to receive unknown corrections at $\ocal(\bar V^2)$. This is
because the $\epsilon + P$ appearing in the definition of $\gamma$
in (\ref{eq:cyclo}) will have a $\bar V^2$ dependence. Our
expression (\ref{eq:timescale}) for $1/\tau$ is correct to order
$\bar V^2$ and in principle we know its full $B$ dependence.
However, we expect the hydrodynamic approximation to fail at
$\ocal(B^3)$. Thus, $1/\tau$ should be trustworthy at order $\ocal
(B^2 \bar V^2)$ but no higher. Problematically, we do not know the
corresponding $\ocal (B^2 \bar V^2)$ correction to $\gamma$. This
uncertainty turns out not to be important for many of our
considerations. In particular, when studying the Nernst effect at
$\rho = 0$ we will find that there is no dependence on $\gamma$.

\section{Impurity relaxation in the truncated M2 brane theory}

We have shown that computing the impurity relaxation timescale
boils down to the two point function of equation
(\ref{eq:tauimp}). In this section we will perform the calculation
for a truncation of the M2 brane CFT, using an adaptation of the
AdS/CFT dictionary for real time two point functions proposed in
\cite{Son:2002sd}.
As noted in the introduction, the truncation of the M2 brane theory is
not consistent, and we make a nonstandard choice of
conformal dimension for one of the scalars.

The CFT at finite temperature with a nonzero charge density and
background magnetic field is dual to a dyonic black hole in
$AdS_4$ \cite{Hartnoll:2007ai}. Such a black hole is a solution to
Einstein-Maxwell theory with a negative cosmological constant.  To
this gravitational theory we would like to add a neutral scalar
field $\phi$ dual to our relevant operator $\ocal$.

From the M2 brane point of view, this Einstein-Maxwell theory
with a neutral scalar should be a sector of eleven dimensional
supergravity compactified on an $S^7$. After the compactification,
we are left with an ${\mathcal N}=8$ supersymmetric gravitational
theory on a four dimensional spacetime with a negative
cosmological constant and infinite towers of Kaluza-Klein states.
The isometry group $SO(8)$ of the $S^7$ determines many features
of the lower dimensional theory.

It is believed to be consistent to truncate this compactification
to the lowest states in the Kaluza-Klein towers. This truncation
is called ${\mathcal N}=8$ gauged supergravity. The bosonic
content of the truncation is the four dimensional graviton, an
$SO(8)$ gauge field, and a scalar and pseudoscalar transforming
under 35 dimensional representations of $SO(8)$. These fields and
their superpartners are thought not to act as sources for the
other excited modes in the Kaluza-Klein towers, and hence the
truncation is called consistent.

There exists a further truncation of ${\mathcal N}=8$ gauged supergravity
to an abelian sector \cite{duffliu}.  Consider the Cartan subalgebra of $so(8)$.
The group $SO(8)$ acts naturally on ${\mathbb R}^8$, and we can think
of the elements of the Cartan subalgebra as generators of rotations in the
12, 34, 56, and 78 planes of $\mathbb{R}^8$.  The action for the abelian truncation
involves only the four $U(1)$ gauge fields corresponding to these four rotations and
the three scalar and pseudoscalar fields neutral under these four $U(1)$ gauge groups.
The full action for this abelian truncation can be found for example in ref.~\cite{Duff}.

To see which scalar fields remain neutral, it is useful to think
of the 35 dimensional representation of $SO(8)$ as the set of
symmetric traceless quadratic polynomials in the coordinates on
${\mathbb R}^8$, $x^i x^j$.  The $U(1)$'s generated by the Cartan
subalgebra act as phase rotations on the complex combinations $z_i
= x^{2i-1} \pm i x^{2i}$ for $i=1,2,3,4$.  Thus the neutral
quadratic scalars must be $|z_i|^2 - |z_{i+1}|^2$. There are three
independent such scalars. By allowing the dual operators $\ocal$
to couple to the impurity potential, we are breaking the $SO(8)$
symmetry of the M2 brane theory. However, the operators are
neutral under the $U(1)$ symmetry for which we have a background
$B$ field and charge density. Indeed, models for `real world'
electronic systems often involve neutral impurities (see for
example \cite{Erginsoy}).

We will make a further truncation of this theory and consider only
the gauge field corresponding to a simultaneous rotation in all
four planes, along with the scalar and pseudoscalar pair
corresponding to just one of these neutral elements of the 35
dimensional representation. As we consider scalar two-point
functions, we will only need the action to quadratic order in the
scalars:
\bea\label{eq:action}
S  & = \displaystyle -\frac{1}{2\kappa_4^2} \int d^4 x \sqrt{-g} &
\Big( R -
\half \left[ (\partial_\mu \phi) (\partial^\mu \phi) + (\partial_\mu \chi) (\partial^\mu \chi) \right]
+ 2 L^{-2} \left[3 +
\half (\phi^2 + \chi^2)  \right] \nonumber \\
 & & - L^2 \left[1 + \half (\phi^2 - \chi^2)  \right] F_{\mu\nu} F^{\mu\nu} + \half L^2 \phi \chi
 \epsilon^{\mu \nu \rho \sigma} F_{\mu\nu} F_{\rho\sigma} \Big) \, .
\eea
Here $\epsilon^{\mu \nu \rho \sigma}$ is the totally antisymmetric
tensor with $\epsilon^{0123} = 1/\sqrt{-g}$, and $L$ is a length
scale which determines the $AdS$ radius. The dyonic black hole is
a solution to the equations of motion provided that $\phi=\chi=0$.
As we can see in (\ref{eq:action}), however, the fact that
$\epsilon^{\mu \nu \rho \sigma} F_{\mu\nu} F_{\rho\sigma}$ does
not vanish for the dyonic background implies that fluctuations in
$\phi$ and $\chi$ source each other. This mixing is why we
keep the two scalar fields. We comment below on
the physical implications of the doubling.

We note that in a charged background, our final truncation to the
two scalars plus the diagonal Maxwell field is not consistent. We
have neglected a quadratic coupling between the neutral scalar
fields and fluctuations of a non-diagonal combination of the four
$U(1)$ gauge fields of the full M2 brane theory \cite{Duff}.
Specifically, we are throwing away the terms
\be
S_G = - \frac{1}{2\kappa_4^2} \int d^4x \sqrt{-g} \left( - L^2
G_{\mu \nu} G^{\mu \nu} - 2 L^2 \phi F_{\mu \nu} G^{\mu \nu} + L^2
\chi \epsilon^{\mu \nu  \rho \sigma} F_{\mu \nu} G_{\rho \sigma}
\right) \,,
\ee
where $G_{\mu \nu} = \delta F_{\mu \nu}^{(1)} + \delta F_{\mu
\nu}^{(2)} - \delta F_{\mu \nu}^{(3)} - \delta F_{\mu \nu}^{(4)}$
is the field strength for a non-diagonal subgroup of $U(1)^4$. We
will continue to ignore this extra field for two reasons. Firstly,
it makes the fluctuation analysis much more complicated. Secondly,
from our condensed matter point of view, we need only one gauge
field to model electricity and magnetism, and it is not clear to
what this other gauge field would correspond. In appendix
\ref{sec:G} we give more details about this coupling.

The black hole has metric
\be\label{eq:metric}
\frac{1}{L^2} ds^2 = \frac{\alpha^2}{z^2} \left[-f(z) dt^2 + dx^2 + dy^2\right] +
\frac{1}{z^2} \frac{dz^2}{f(z)} \,,
\ee
and carries both electric and magnetic charge
\be\label{eq:Ffield}
F = h \alpha^2 dx \wedge dy + q \alpha dz \wedge dt \,,
\ee
where $q, h$ and $\alpha$ are constants. The function
\be
f(z) = 1 + (h^2 + q^2) z^4 - (1 + h^2 + q^2) z^3 \,.
\ee
The Hawking temperature, dual magnetic field and charge density of
this black hole are \cite{Hartnoll:2007ai}
\be
B = h \alpha^2 \,, \qquad \rho = - q \alpha^2 \sigma_0
\,, \qquad T = \frac{\alpha (3-h^2-q^2)}{4 \pi} \,.
\ee
Note that $q^2+h^2 \leq 3$, whereas $\rho$ and $B$ are unbounded.
We have expressed the charge density in terms of the the
conductivity of the dual CFT at $\rho=B=0$,
\be
\sigma_0 = 2 L^2/\kappa_4^2 = \sqrt{2} N^{3/2}/6\pi \,.
\ee
In this last equality, we re-expressed the dimensionless
$L^2/\kappa_4^2$ in terms of the number $N$ of coincident M2
branes \cite{MAGOO}. These relations can be inverted to give
$\alpha(T,\rho,B)$ as the solution of
\be
3 \alpha^4 - 4 \pi T \alpha^3 = B^2 + \rho^2/\sigma_0^2 \,.
\ee

We would now like to consider small fluctuations of the scalar
fields $\phi$ and $\chi$ about this black hole background. It is
useful to diagonalise the quadratic Lagrangian (\ref{eq:action})
first by defining
\be\label{eq:psi}
\psi_+ = \frac{h \phi + q \chi}{\sqrt{q^2+h^2}} \qquad \mbox{and} \qquad
\psi_- = \frac{q \phi - h \chi} {\sqrt{q^2+h^2}} \ .
\ee
The two fields will be dual to two relevant operators
$\ocal_\pm$.
For our dyonic black hole background
\be
L^4 F_{\mu\nu} F^{\mu\nu} = 2 z^4 (h^2-q^2) \,, \qquad \half L^4
\epsilon^{\mu \nu \rho \sigma} F_{\mu\nu} F_{\rho\sigma}  = - 4
z^4 h q \,.
\ee
Substituting into the action (\ref{eq:action}) one obtains the
following decoupled actions for the scalar fields
\be
S_\pm = - \frac{1}{2\kappa_4^2} \int d^4 x \sqrt{-g} \left( -
\frac{1}{2} (\partial_\mu \psi_\pm) (\partial^\mu \psi_\pm) +
\frac{1}{L^2} \left[1 \mp z^4 (q^2+h^2) \right] \psi_\pm^2
\right) \,.
\ee
The equations of motion are therefore
\be
\Box \psi_\pm = (m^2 \pm 2 z^4 (q^2 + h^2)/L^2 )\psi_\pm \, ,
\ee
where $m^2 L^2 = -2$. Note that this mass is above the
Breitenlohner-Freedman bound $m^2 L^2 \geq -9/4$. Depending on the
sign of $\pm$, the electromagnetic field provides a potential for
the scalar that either pushes the scalar away from or towards the
horizon. We will assume that the scalar field has the following
dependence on $x$, $y$, and $t$: $\phi \sim e^{i k x - i
\omega t}$.  Thus, the equation of motion becomes
\be\label{eq:phieqn}
z^4 \left( \frac{f}{z^2} \psi_\pm' \right)' - \fk^2  z^2 \psi_\pm
+ \fw^2 \frac{z^2}{f} \psi_\pm - m^2 L^2 \psi_\pm \mp 2 z^4 (q^2 +
h^2) \psi_\pm= 0 \ ,
\ee
where we have defined the dimensionless frequency and wavevector
$\fw \equiv \omega / \alpha$ and $\fk
\equiv k / \alpha$, and also $f' = df / dz$. Near the boundary, $z=0$, of this
asymptotically $AdS_4$ geometry, the scalar has the usual behavior
\be
\psi_\pm = z^{3-\Delta} ( A(\w,k) + {\cal O}(z^2) ) + z^\Delta (B(\w,k) +
{\cal O}(z^2))\,,
\ee
where $\Delta (\Delta-3) = m^2 L^2$.

In the standard M2 brane theory, representation theory of
$SO(8)$ and the superconformal symmetry guarantees
that $\phi$ has conformal dimension $\Delta_\ocal=1$
while the pseudoscalar $\chi$ has dimension $\Delta_\ocal=2$
\cite{MAGOO,Aharony:1998rm,Minwalla:1998rp, Halyo:1998mc}.
Introducing nonzero $B$ and $\rho$, as we have seen, mixes $\phi$
and $\chi$.  We speculate that in the dyonic black hole
background, it is the linear combinations $\psi_\pm$ which have
definite conformal dimension rather than $\phi$ and $\chi$
independently. We shall avoid these subtleties by taking both
$\phi$ and $\chi$ to have dimension $\Delta_\ocal=1$.  The
standard M2 brane theory is related to this choice of conformal
dimension by a deformation by a double trace operator
\cite{Witten:2001ua}. This choice has the added benefit of
providing a second scalar to analyze with conformal dimension one,
and therefore relevant.

The dimensions $\Delta_\ocal=1$ and $\Delta_\ocal=2$ both
correspond to the mass $m^2 L^2 = -2$. The fact that there are two
possible conformal dimensions associated with the same mass has
important consequences here. Having chosen the smaller conformal
dimension, it is the faster falloff that is dual to a source for
$\ocal$ in the boundary theory \cite{Klebanov:1999tb}
\be
J_\ocal = \alpha^{3-\Delta_\ocal} A(\w,k)\,,
\ee
whereas the slower falloff gives the expectation value
\be
\langle \ocal \rangle =
\frac{\sigma_0}{4} \alpha^{\Delta_\ocal} (2 \Delta_\ocal - 3) B(\w,k) \,.
\ee
From these expressions one can read off the Green's function
\be
\langle \ocal \rangle = G_{\ocal \ocal}^R(\w, k) J_\ocal \,.
\ee
Thus
\be
G_{\ocal \ocal}^R(\w, k) = - \frac{\sigma_0}{4\alpha}
\frac{B(\w,k)}{A(\w,k)} \ .
\label{eq:ourprescription}
\ee
The standard procedure for scalars of mass $m^2 L^2 = -2$ would
have yielded $G_{\ocal \ocal}^R \sim A/B$.  Because our operators
have the smaller conformal dimension,
we get the result (\ref{eq:ourprescription}).

Our job is to determine the dimensionless function $F$ appearing
in the scaling relation (\ref{eq:tscaling}). We will need to
compute two scaling functions $F_\pm$, one for each scalar field
$\psi_\pm$. An observation that simplifies our task is that the
differential equation (\ref{eq:phieqn}) depends on $B$ and $\rho$
only in the combination $h^2+q^2$, or equivalently $B^2 + \rho^2/
\sigma_0^2$. Thus, $F$ is a function of $B^2 + \rho^2/\sigma_0^2$
only\footnote{%
Amusingly, this dependence only on the combination $h^2+q^2$
remains true in the full M2 brane theory, in which one keeps the
second gauge field to which $\phi$ and $\chi$ couple. See appendix
\ref{sec:G}.}. Using the fact that for the dyonic black hole
\cite{Hartnoll:2007ai, HartnollHerzog}
\be
\epsilon = \frac{\sigma_0 \alpha^3}{2} (1+h^2+q^2) \,, \qquad
P = \frac{\epsilon}{2} \,, \qquad s = \pi \sigma_0 \alpha^2 \,,
\ee
and that for the M2 theory (\ref{eq:tauimp}) becomes
\be\label{eq:taum2}
\fbox{$\displaystyle
\frac{1}{\tau_\text{imp}} =
\frac{\bar V^2}{T} F\left(\frac{1}{T^2} \sqrt{B^2 + \rho^2/\sigma_0^2}\right)
\,,
$}
\ee
the expressions in the preceding paragraphs imply that
\be\label{eq:F1}
\fbox{$\displaystyle
F\left(\frac{1}{T^2} \sqrt{B^2 + \rho^2/\sigma_0^2}\right) =
-\frac{s T}{16 \pi^2 (\epsilon + P)}
\lim_{\fw \to 0} \int d\fk \,
\fk^3 \, \text{Im} \frac{B(\fw,\fk)}{\fw A(\fw,\fk)} \,.
$}
\ee
It is interesting to recall from \cite{HartnollHerzog} that the dc
conductivity for the M2 brane, with nonvanishing $B$ and $\rho$ is
given by
\be
\sigma = \frac{(sT)^2}{(\epsilon+P)^2} \sigma_0 \,,
\ee
allowing us to rewrite (\ref{eq:F1}) as
\be
F\left(\frac{1}{T^2} \sqrt{B^2 + \rho^2/\sigma_0^2}\right) =
-\frac{1}{16 \pi^2}
\sqrt{\frac{\sigma}{\sigma_0}}
\lim_{\fw \to 0} \int d\fk \, \fk^3 \, \text{Im} \frac{B(\fw,\fk)}{\fw A(\fw,\fk)} \,.
\ee
The physical meaning of this formula is not clear, because in
general there's no particular reason for the impurity relaxation
timescale to be related to the electrical conductivity of the CFT
in the absence of impurities. It remains for us to solve for
$A(\fw,\fk)$ and $B(\fw,\fk)$ numerically.

\subsection{Comment on discrete symmetries}

Before turning to computations, we should comment that
we have found two relevant operators $\ocal_\pm$, both with
dimension $\Delta_{\ocal_\pm} = 1$. Both operators preserve the
global $U(1)$ symmetry and in principle, they could both be
sourced by impurities.  However, these operators transform
differently under the discrete symmetries charge conjugation $\CD$, parity $\PD$,
and time reversal $\TD$.

Usually we make a sharp distinction between the way operators
transform and the way states transform.  The situation here is
complicated by the fact that our fields $\psi_\pm$ dual to the
${\mathcal O}_\pm$, and hence the ${\mathcal O}_\pm$ themselves,
are defined in (\ref{eq:psi}) in terms of the state and in
particular the choice of background magnetic field $B$ and charge
density $\rho$. In studying $\psi_\pm$, when we act with $\CD$,
$\PD$, and $\TD$, we will act on both the operator and the
underlying state.

From the gauged ${\mathcal N}=8$ supergravity construction we know
that $\phi$ and $\chi$ are both real fields and thus do not transform
under $\CD$.  At the same time $\phi$ is a scalar while $\chi$ is a pseudoscalar.
Thus we have $\PD(\phi) = \phi$ and $\PD(\chi) = -\chi$.  From the CPT theorem,
we must have under time reversal that $\TD(\phi) = \phi$ while $\TD(\chi)=-\chi$.

The transformation properties of the electromagnetic field under $\CD$, $\PD$, and $\TD$
are well known.  In our case, the strength of the
electric field of the black hole, which is dual to the charge density
$\rho$ on the boundary, is characterized by the value of $q$.
The magnetic field strength $B$ is fixed by the value of $h$.
Under parity, we know the electric field is a vector while the magnetic
field is a pseudovector.  Thus we have $\PD(q) = -q$ while $\PD(h) = h$.
$\CD$ flips the sign of both $q$ and $h$.
Finally, the magnetic field famously breaks time reversal symmetry, and
so $\TD(h) = -h$ while $\TD(q)=q$.

Assembling these facts about $q$, $h$, $\phi$ and $\chi$ together we find that
\be
\begin{array}{lll}
\CD(\psi_+) = - \psi_+ \ ,& \PD(\psi_+) = \psi_+ \ ,& \TD(\psi_+) = - \psi_+\ , \\
\CD(\psi_-) = - \psi_- \ , & \PD(\psi_-) = -\psi_- \ ,& \TD(\psi_-) =  \psi_- \ .
\end{array}
\ee
The subscript of $\psi_\pm$ thus corresponds to the eigenvalue of
the field under $\PD$.  In adding impurities, we are faced with a
choice: we can let the impurities break $\TD$ but not $\PD$ or
$\PD$ but not $\TD$. In the next section, we will see the behavior
of the scattering time $\tau_{\text{imp}}$ for the ${\mathcal
O}_+$ impurities is markedly different from the ${\mathcal O}_-$
impurities.

Note that ${\mathcal O}_+$ transforms the same way under the
discrete symmetries as the magnetic field while ${\mathcal O}_-$
transforms as the electric field.  It is thus tempting to call the
${\mathcal O}_+$ impurities magnetic and the ${\mathcal O}_-$
impurities electric. The underlying scalar fields $\phi$ and
$\chi$ are of course neutral. The nontrivial transformation of
$\psi_\pm$ under $\CD$ comes from the dressing by the dyonic black
hole background, or equivalently, the presence of magnetic field
and charge density in the field theory.

\section{Numerical results for $1/\tau_\text{imp}$}

The prescription for computing the retarded Green's function
\cite{Son:2002sd} is that we solve the equation for our scalar field in the
black hole background (\ref{eq:phieqn}) with ingoing boundary
conditions at the black hole horizon $z=1$,
\be\label{eq:ingoing}
\psi \sim (1-z)^{i \fw / (h^2 +q^2-3)} \ .
\ee
It is convenient to impose these boundary conditions by defining a
new function $S(z)$ such that
\be
\psi(z) \equiv \exp \left(i \fw \int_0^z \frac{x^2}{f(x)} dx\right) S(z)\
.
\ee
Thus $S(z)$ is nicely behaved at the horizon and we can impose
$S(1) = 1$. It is straightforward to integrate the equation for
$S(z)$ numerically from the horizon to the boundary at $z=0$. Near
the boundary one can read off the value of the ratio $B/A$ that we
need to compute the scaling function $F$.

Given $B/A$, we then need to compute the integral (\ref{eq:F1})
over $k$.  This integral may not converge, and indeed we discuss
below a large $\rho$ and $B$ regime where it does not for the
$\psi_-$ scalar field.  However, the divergence comes from bad
behavior at finite $k$.  At large $k$ we expect the imaginary part
of the Green's function to be exponentially damped.  Appendix
\ref{sec:WKB} presents a WKB analysis that demonstrates this
damping.  Qualitatively, we may understand the behavior from the
bulk as follows.  At large $k$, the Green's function is dominated
by a space-like geodesic which stays close to the boundary of the
space-time and far from the black hole horizon; it is the horizon
that is responsible for producing dissipative effects and a
nonzero contribution to the imaginary part of the Green's
function.

\begin{figure}[h]
\begin{center}
a) \epsfig{file=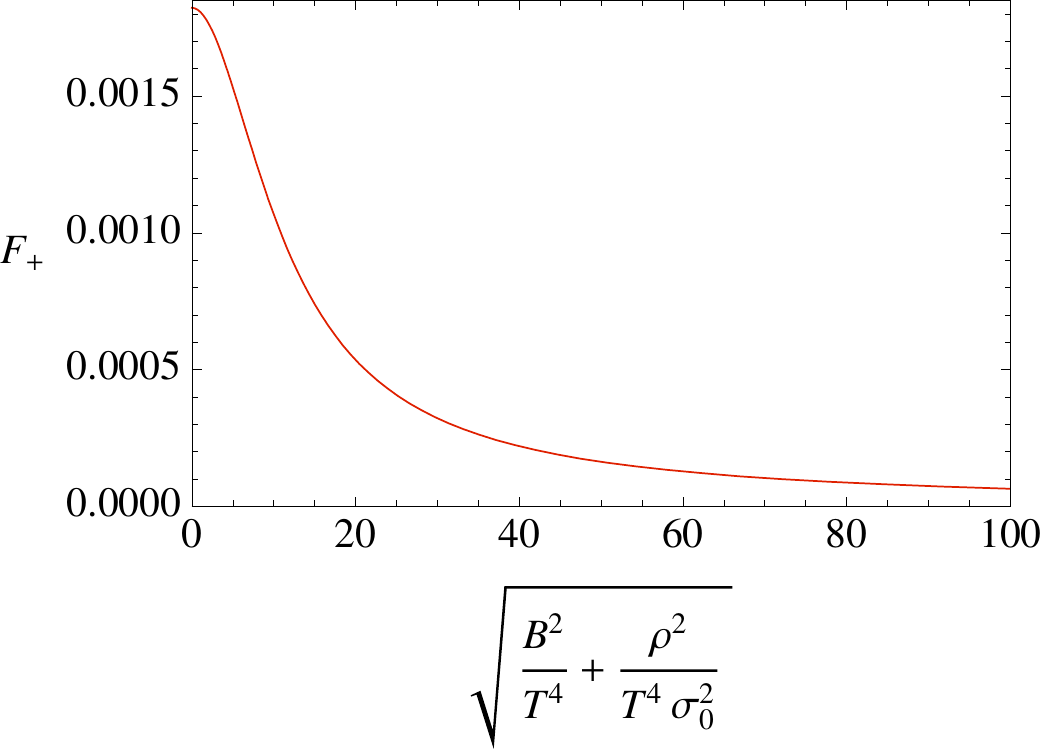,width=7.19cm}%
b) \epsfig{file=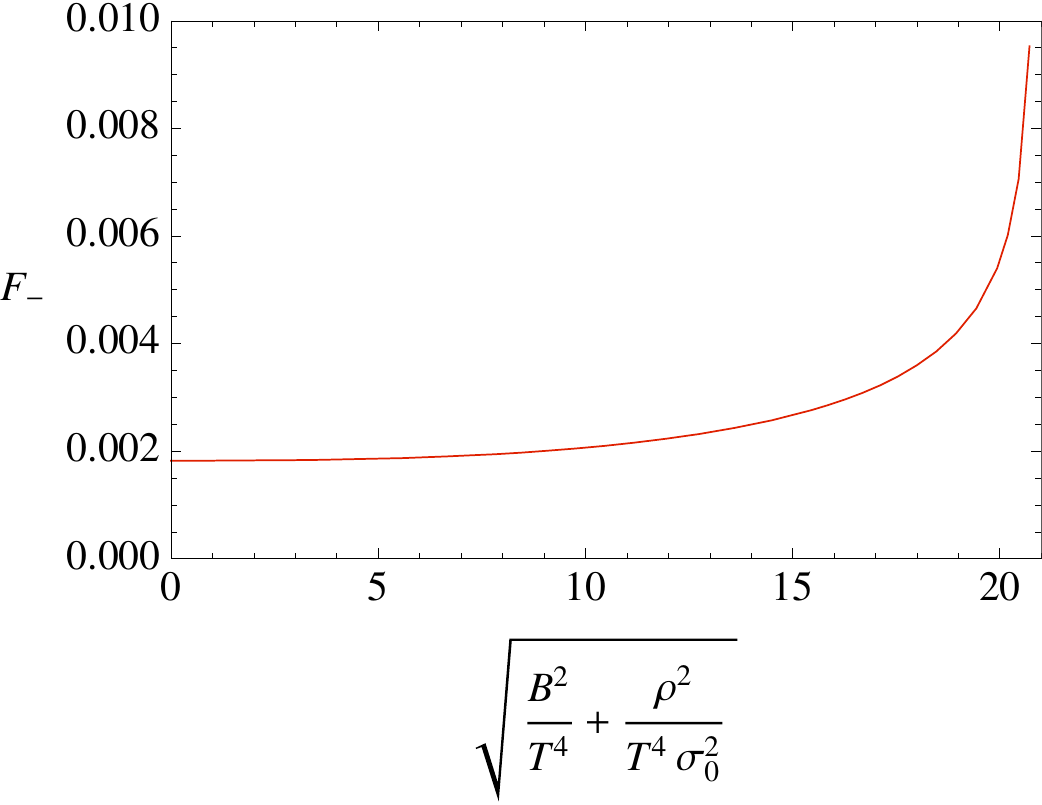,width=6.81cm}
\end{center}
\caption{The function $F_\pm$ for the M2 brane theory
for a) the scalar field $\psi_+$ and b) $\psi_-$.}
\label{fig:norho}
\end{figure}

\begin{figure}[h]
\begin{center}
a) \epsfig{file=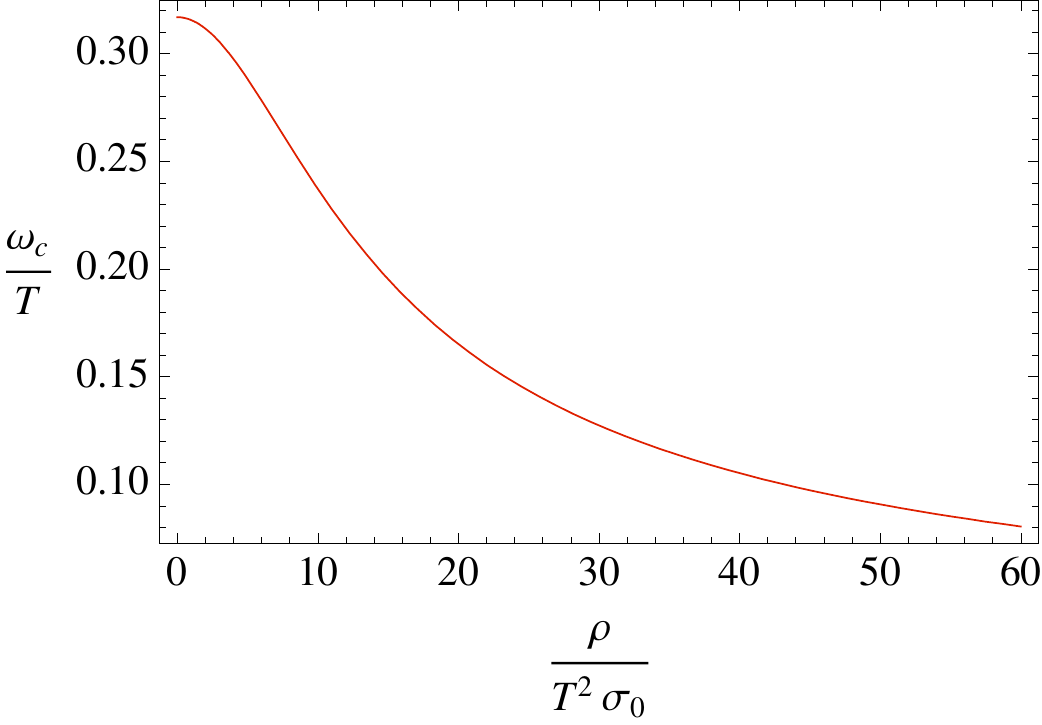,width=7.cm}
b) \epsfig{file=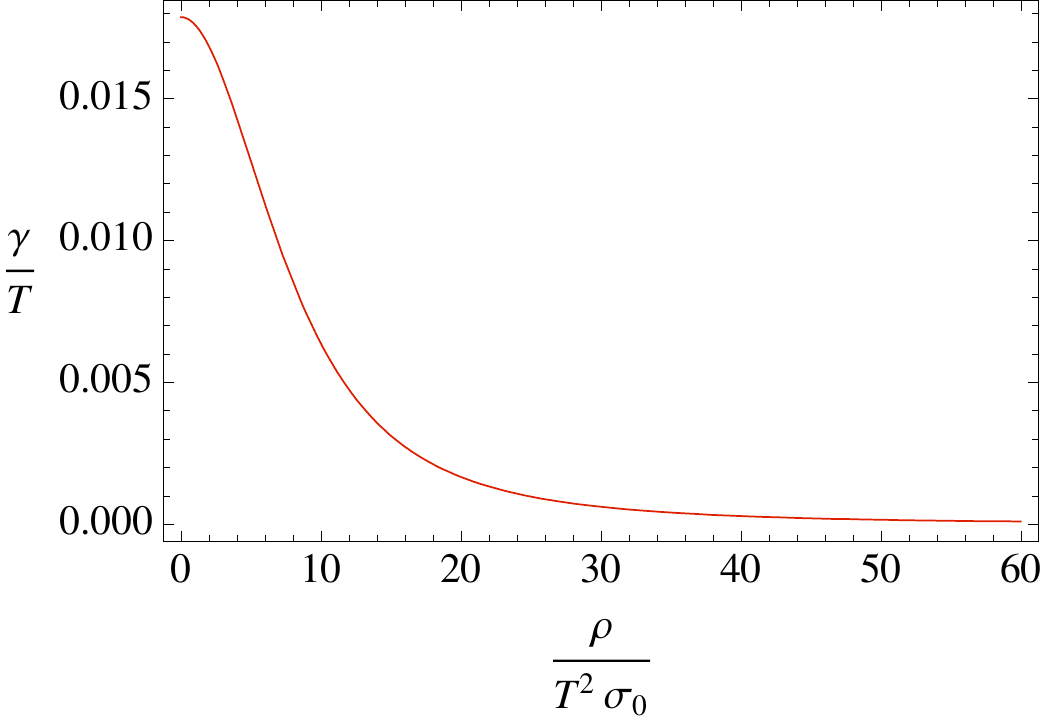,width=7.cm}
\end{center}
\caption{a) The real part of the cyclotron frequency and b) momentum relaxation due to the cyclotron resonance at  $B = T^2$.}
\label{fig:gamma}
\end{figure}

We begin by considering the two point function corresponding to
the scalar field $\psi_+$, that is, magnetic impurities. The
result for $F_+$ is plotted in figure \ref{fig:norho}a. As the
combination $B^2 + \rho^2/\sigma_0^2$ is increased, the inverse
scattering time becomes smaller and smaller. There is also a
competing physical process that we described in section
\ref{sec:mag} above: the cyclotron resonance. If we keep $B/T^2$ small
($h \ll 1$), then we can trust our hydrodynamic approximation of
the location of the cyclotron pole at $\omega_c - i \gamma$, given
in (\ref{eq:cyclo}). In figure \ref{fig:gamma}, we plot the
dimensionless $\gamma/T$ against $\rho/T^2 \sigma_0$ for an
appropriately small $B/T^2 =1$. We see that at large $\rho$, the
damping due to the cyclotron resonance also gets small while the
frequency changes more slowly. The reduction in damping both from
impurities and the $B$ field at large $\rho$ may make the
hydrodynamic cyclotron mode easier to observe experimentally than
the results in \cite{Hartnoll:2007ih} suggested, as there the
dependence of $\tau_\text{imp}$ on $B$ and $\rho$ was not
considered.

We also would like to consider the scaling function $F_-$ for the
$\psi_-$ scalar fields.  The result is plotted in figure
\ref{fig:norho}b, and is somewhat dramatic.  We define
the related dimensionless quantities
\be
Q \equiv \sqrt{h^2+q^2} \qquad \mbox{and} \qquad {\mathcal Q}
\equiv \frac{1}{T^2} \sqrt{B^2 + \rho^2/\sigma_0^2} \ .\ee We find
that $F_-$ increases monotonically as a function of $Q$ or
${\mathcal Q}$ and at a finite value $Q_*=0.76654$, which
corresponds to ${\mathcal Q}_*=  20.80 $, $1/\tau_{\rm{imp}}$
diverges. This divergence reveals some interesting physics in the
`pure' CFT which we will return to shortly. Note that we cannot
trust our memory function formalism if $1/\tau_{\rm{imp}}$ gets
too large. Higher order and nonperturbative effects are important
in this regime.

From a gravitational standpoint, we may attempt a qualitative
understanding of the decrease (increase) of $1/\tau_{\rm{imp}}$
with $\rho$ and $B$ for the $\psi_+$ ($\psi_-$) field. From
(\ref{eq:phieqn}), the scalar $\psi_\pm$ experiences a potential
of the form $\pm 2 z^4 (h^2+q^2)$. Thus for $\psi_+$, the
potential repels the scalar from the black hole horizon while for
$\psi_-$, the potential has the opposite effect. The black hole
horizon is responsible for dissipation in the system, and we
expect the more the probability distribution of the scalar is
peaked near the horizon, the larger the imaginary part of the
scalar Green's function will be and hence the larger
$1/\tau_{\rm{imp}}$ will be.  The decrease in $F_\pm$ with $B$ and
$\rho$ for $\psi_+$ and the corresponding increase for $\psi_-$ in
figures \ref{fig:norho}a and \ref{fig:norho}b are thus
qualitatively understood.

\subsection{A black hole instability}

\begin{figure}[h]
\begin{center}
a) \epsfig{file=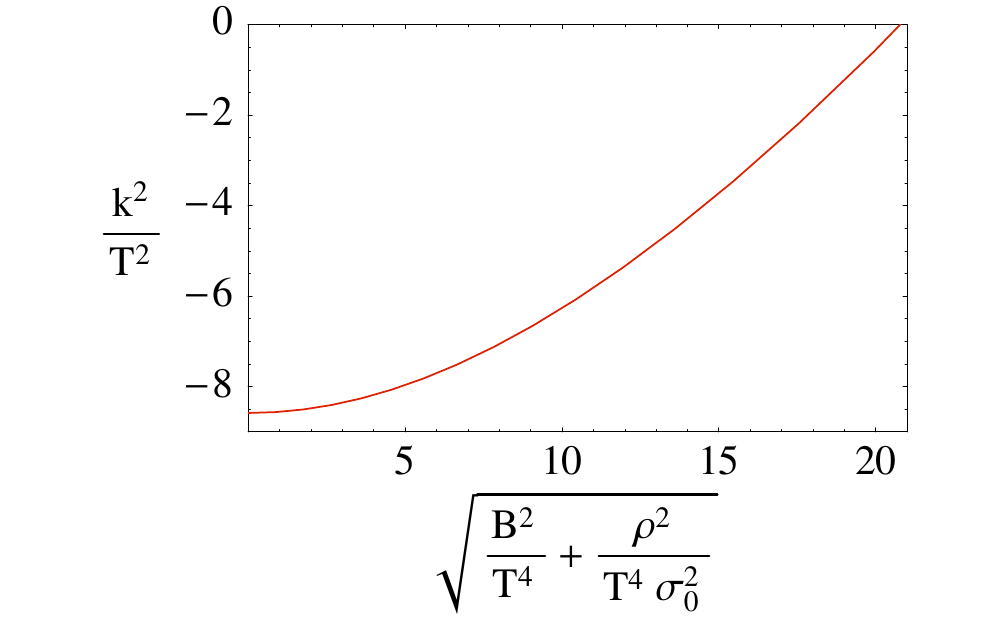,width=7.cm} b)
\epsfig{file=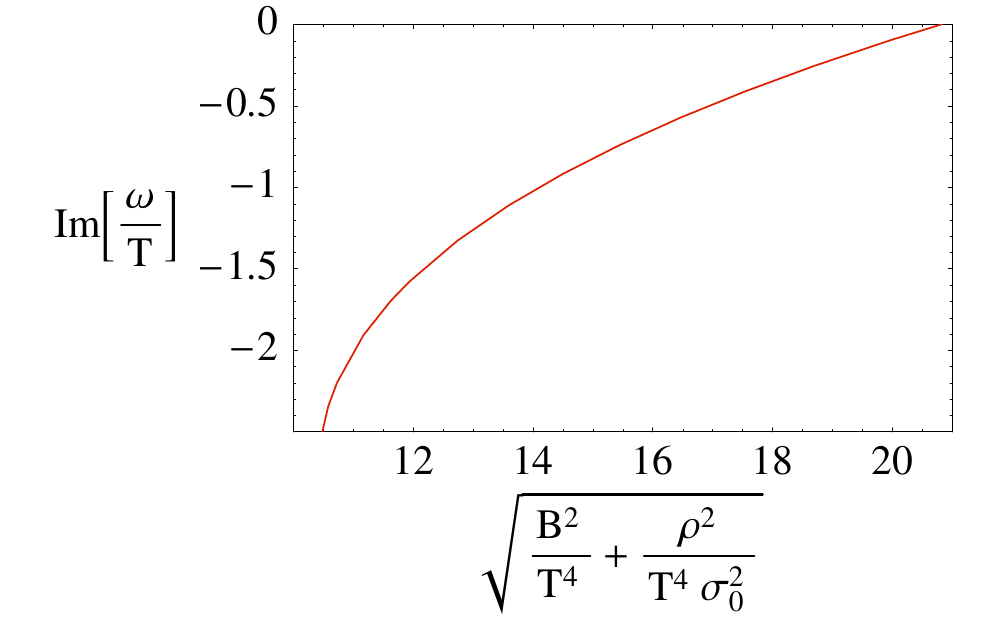,width=7.cm}
\end{center}
\caption{The location of the smallest pole in $G^R_{\ocal_- \ocal_-}$
as a function of $B$ and $\rho$: a) The pole in the complex $k$
plane for $\omega = 0$.  Note $k^2$ is real.  b) The pole in the
complex $\omega$ plane for $k=0$.  Note $\mbox{Re}(\omega)=0$. }
\label{fig:unstablepoles}
\end{figure}

The divergence in $F_-$ is caused by an underlying instability in
the truncated M2 brane theory at large ${\mathcal Q}$. Numerically
we have found that the divergence in $F_-$ appears to be caused by
a diffusion type pole in the scalar Green's function moving onto
the real $k$ axis.
 Near the critical value $Q_*$, the pole has
the approximate form
\be\label{eq:trial}
G_{\ocal \ocal}^R \sim \frac{1}{C(h,q) - \fk^2 + 6 i \fw/5+\ldots}
\ .
\ee
The ellipsis denotes higher order terms in $\omega$ and $k$
and $C(h,q) =1.3911 (Q -Q_*) $.
In figures \ref{fig:unstablepoles}a
and b we have plotted the location of the pole as a function of
${\mathcal Q}$. Note that in both plots, for $k^2$ and $\text{Im} \,
\w$, the curves reach the real axis at ${\mathcal Q}_* = 20.80$. At
this point, two changes occur simultaneously. First,
$1/\tau_\text{imp}$ diverges because, at $\w = 0$, there is a
(double) pole on the real $k$ axis that is integrated over in
(\ref{eq:F1}). Second, at $k=0$, the imaginary part of $\omega$
switches sign leading to exponential growth in the scalar mode at
long times.

This exponential growth in time is indicative of an underlying
instability in the CFT when $Q>Q_*$.
 Indeed, the
retarded Green's function cannot have poles in the upper half $\w$
plane, indicating that for $Q >Q_*$ we are no longer in the
correct vacuum. To evaluate the impurity time for $Q > Q_*$, we
would need to find a different stable supergravity background in
which the $\psi_-$ field was at a local minimum. This instability
illustrates the dangers in this particular black hole background
of working in a consistent truncation in which we set $\psi_-=0$;
had we ignored fluctuations in $\psi_-$, we would have missed this
instability.

This instability in the theory is interesting from a purely
gravitational perspective. First note that poles in the Green's
function (\ref{eq:ourprescription}) occur when the denominator
$A(\w,k)$ vanishes. This vanishing means there exist on shell
modes with the behaviour $\psi_\pm = z + {\mathcal{O}}(z^3)$ near
the conformal boundary $z=0$. Due to the nonstandard boundary
conditions for the $\psi_\pm$ fields, these modes are physical.
For the critical value $Q = Q_*$, the physical mode is static,
with $\w = 0$. This `threshold' mode separates oscillating modes,
with $Q < Q_*$ and $\text{Im}\, \w < 0$, from exponentially
growing modes, with $Q > Q_*$ and $\text{Im}\, \w > 0$.  For $Q >
Q_*$ there is an on shell classical instability of the black hole.

A classical instability of the full finite temperature M2 brane
was discovered in \cite{Gubser:2000mm, Gubser:2000ec}. To see that
instability, it is necessary to retain more than one Maxwell field
in the bulk, whereas we have truncated to a diagonal combination.
Converting to our variables, the Gubser-Mitra instability occurs
with $h=0$ and at $q_{GM} = 1$, corresponding to the value
$\rho_{GM}/\sigma_0 T^2 \approx 39.5$. This charge density is
larger than necessary for the instability we have described. Given
that the two instabilities occur in different theories, one might
be cautious about attaching much meaning to the relative values of
the critical charges\footnote{In fact, the analysis in
\cite{Gubser:2000mm, Gubser:2000ec} also employs an inconsistent
truncation. Although that work does include the gauge field $G$,
it does not include the pseudoscalar $\chi$, which is sourced by
$G$. See appendix \ref{sec:G}. 
For an investigation of threshold modes $\omega=k=0$, 
this $\chi G_{xy}$ coupling will vanish and so should not affect the location 
of the Gubser-Mitra instability.
Note also, \cite{Gubser:2000mm,
Gubser:2000ec} work with conformal dimension $\Delta = 2$ for the
operator dual to $\phi$, so their dual theory is also related to
the standard M2 brane theory by a double trace deformation.}.

The classical Gubser-Mitra instability occurs at precisely the
same value of the charge density at which the charged black hole
with four $U(1)$ gauge fields
becomes thermodynamically unstable.
 This coincidence led Gubser and Mitra to the `correlated
stability' conjecture, that for translationally invariant
horizons, thermodynamic and classical dynamic instabilities should
always coincide.

In contrast, our black hole with only one $U(1)$ gauge field is
always thermodynamically stable
\cite{Hartnoll:2007ai}; it only becomes thermodynamically unstable
when embedded into the full M2 brane theory, with four independent
charges.
Given this dynamical instability without a corresponding
thermodynamic instability, we have a counterexample to
the correlated stability conjecture. The physics underlying our
counterexample appears to be very similar to the counterexamples
discussed in \cite{FGM}. The mismatch between thermodynamic and
classical instability is possible because the instability is due
to scalar fields that are not associated to conserved charges. The
instability indicates a phase transition to an ordered phase,
characterised by a condensate for the operator $\ocal_-$ dual to
the field $\psi_-$. As noted in \cite{FGM}, it seems appropriate
to widen the notion of thermodynamic instability of black holes in
these cases to allow for variations of the value of the scalar
field at the conformal boundary.  Because this phase transition is associated
with breaking an underlying ${\mathbb Z}_2$ symmetry, $\psi_- \to - \psi_-$,
of the truncated supergravity action, the transition is likely to be second order.

\section{The Nernst effect in the truncated M2 brane theory}

Place a system in a thermal gradient and a perpendicular magnetic
field. The Nernst effect is the observation of an electric field
that is created orthogonal to both the thermal gradient and the
background magnetic field. The Nernst coefficient is the electric
field generated per unit of thermal gradient and magnetic field
\be
N = \frac{E}{B \nabla T} \,.
\ee

Recent experimental interest in the Nernst effect in
superconductors was sparked by the observation by Ong et al.\ of
an anomalously large Nernst effect in the pseudogap region of the
high-T$_c$ superconductors \cite{ong1}. This observation was
interpreted as signalling the presence of vortices, and hence the
persistence of a phase disordered Cooper pair condensate, above
the transition temperature. The Nernst effect has since been used
to probe the presence of long lived Cooper pair fluctuations in
conventional superconductors \cite{behnia1} and also to capture
the proximity of a Mott insulating phase in certain organic
superconductors \cite{ardavan}.

The Nernst effect depends strongly on the impurity relaxation
time. Therefore, the $B$ and $\rho$ dependence of
$\tau_\text{imp}$ that we explored above may have measurable
consequences for observations of the Nernst effect in various
types of superconductors. In this section, we explore the
dependence of the Nernst effect on $\tau_\text{imp}$. We argue
that there are qualitative similarities between aspects of the $T$
and $B$ dependence of the Nernst effect observed in organic
superconductors \cite{ardavan} and the Nernst effect we predict at
$\rho=0$. In the underdoped cuprate superconductors
\cite{ong1, ong2}, for a range of typical values of the doping,
we find that the relatively large size of $\rho$ compared with
experimentally accessible values of $B$ make $\tau_\text{imp}$
relatively insensitive to $B$. The exception to this last
statement is very close to the insulating state at doping $x_I =
1/8$, where $\rho = 0$ is again appropriate.

A hydrodynamic approach to the Nernst effect was developed in
ref.~\cite{Hartnoll:2007ih}. That work led to the following formula
for the Nernst coefficient in a relativistic theory with `speed of
light' $v$. For the next couple of formulae only we shall not set
$v=1$, as we have been doing in this paper, and shall also show
explicitly the unit electric charge $e$.
It was found that
\be\label{eq:nernsthydro}
N = \frac{v^2}{T}
\left(\frac{1/\tau_\text{imp}}{(\w_c^2/\gamma + 1/\tau_\text{imp})^2 + \w_c^2}
\right) \,.
\ee
Restoring the $v$ and $e$ dependence of the cyclotron pole, we have
\be
\w_c = \frac{e B \rho v^2}{\epsilon + P} \qquad \mbox{and}  \qquad \gamma =
\frac{\sigma B^2 v^2}{\epsilon + P} \,.
\ee

In ref.~\cite{Hartnoll:2007ih} the hydrodynamic expression
(\ref{eq:nernsthydro}) was used to reproduce some features of the
experimentally observed Nernst effect in the cuprates. However, in
the absence of a microscopic theory, the simplifying assumption
was made that the functions of state, the conductivity $\sigma$
and the impurity scattering time $\tau_\text{imp}$ did not depend
significantly on the dimensionless ratios $\rho/T^2$ and $B/T^2$.
For the M2 brane theory, $\sigma(\rho/T^2,B/T^2)$ was computed in
ref.~\cite{HartnollHerzog}, and in this paper we have computed
$\tau_\text{imp}$. Thus we have all the ingredients necessary for
an exact computation of the Nernst coefficient at a quantum
critical point.

The Nernst effect is a direct probe of the impurity relaxation
time. We see in (\ref{eq:nernsthydro}) that a nonzero
$\tau_\text{imp}$ is necessary for the Nernst coefficient to be
finite. This connection becomes particularly striking in the limit
of vanishing charge density, $\rho = 0$, where
(\ref{eq:nernsthydro}) becomes simply
\be\label{eq:Nnorho}
N = \frac{v^2 \tau_\text{imp}}{T} = \frac{v^2}{\bar V^2
F(B/T^2)}\,,
\ee
where in the second step we used (\ref{eq:taum2}) for the M2 brane
theory. One nice feature of this formula is that the dependence on
$\gamma$ has dropped out, and with it the uncertainties of order
$\bar V^2 B^2$ that we discussed at the end of section
\ref{sec:mag}.

\begin{figure}[h]
\begin{center}
a)\epsfig{file=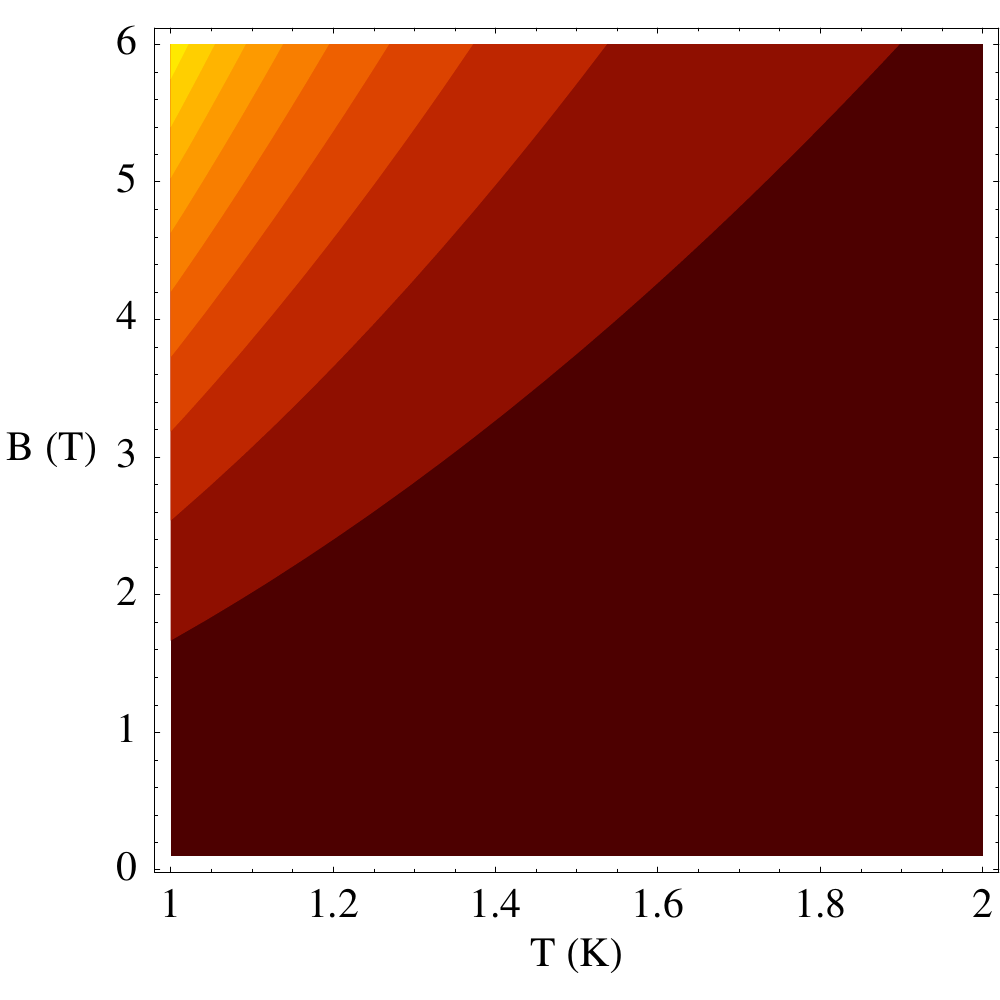,width=6cm}
b)\epsfig{file=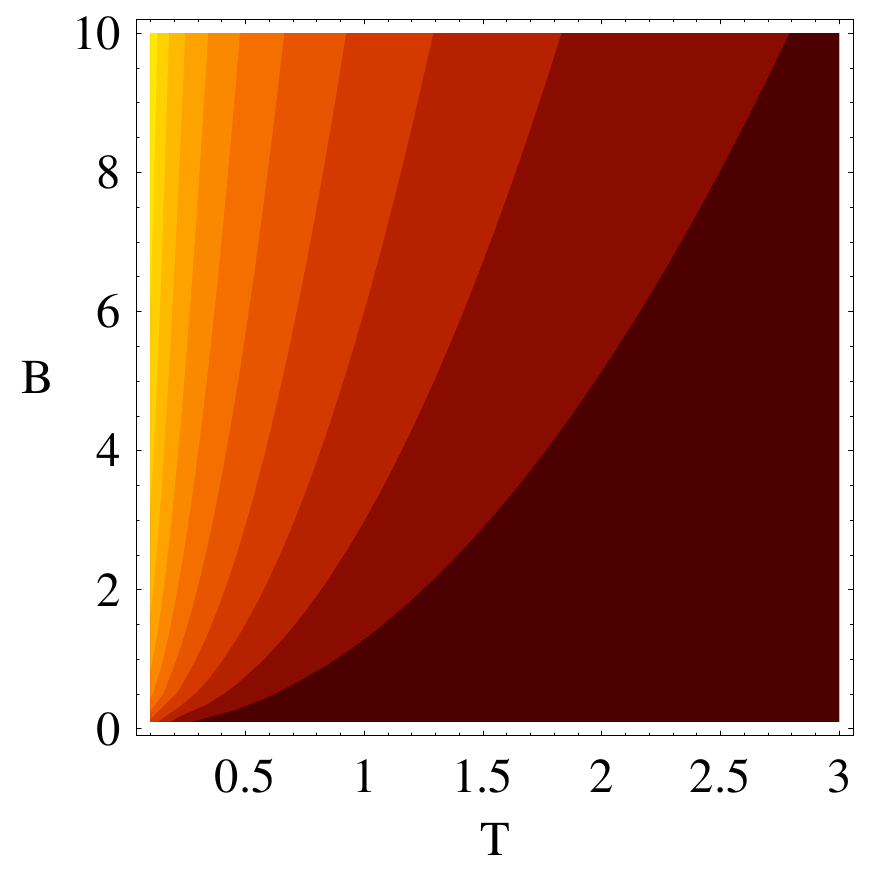,width=6cm}
\end{center}
\caption{(a) The Nernst signal $N$ for the truncated M2 brane
theory as a function of $B$ and $T$, with vanishing charge density
$\rho = 0$. The impurity potential is coupled to $\ocal_+$.
Lighter denotes a larger Nernst coefficient. (b) The Nernst signal
in the vicinity of a general quantum critical point with $\rho =
0$ and $\Delta_\ocal = 1$. The shading is logarithmically spaced.}
\label{fig:contournorho}
\end{figure}

Using our numerical results from the previous section, it is
straightforward to plot the Nernst coefficient for the truncated
M2 brane theory as a function of $B$ and $T$, which is how the
experimental data is often presented \cite{ardavan,ong2,behnia2}.
In figure
\ref{fig:contournorho}a we plot the Nernst coefficient in the case
of vanishing charge density (\ref{eq:Nnorho}). We have taken the
impurity potential to couple to the $\ocal_+$ operator. If we had
taken the $\ocal_-$ operator, the dark and light regions of the
plot would have been interchanged. For ease of comparison with
experiments, we should consider plotting in SI units. In order to
convert $B/T^2$ from our `natural units', with $v=k_B=\hbar=e=1$,
one must first introduce the velocity $v = \Phi_v {\rm m/s}$, and
then note that
\be
\frac{\text{tesla}}{\text{(kelvin)}^2} = 8.8 \times 10^{-8} \Phi_v^2
\frac{k_B^2}{e \hbar v^2} \,.
\ee
We now need to specify the number $\Phi_v$. A convenient choice is
$\Phi_v = 3400$, as this makes the conversion factor $8.8 \times
10^{-8} \Phi_v^2 = 1$. Curiously, this appears to be in a
physically reasonable ballpark, at least for the cuprate
superconductors. This value corresponds to $\hbar v = 22\,
\text{meV} \, \ang$, whereas the value estimated in \cite{Hartnoll:2007ih} for
underdoped LSCO was $47\, \text{meV}\, \ang$ and the characteristic
velocity observed in YBCO was $35\, \text{meV} \, \ang$
\cite{balatsky}.  While all plots in this section are in natural units, in
figure \ref{fig:contournorho}a, for concreteness only, we have
assumed $\hbar v = 22 \, \text{meV} \, \ang$ and put kelvin and
tesla on the axis labels.

In figure \ref{fig:contournorho}a, the range of temperatures and
magnetic fields has been constrained by the validity of the
hydrodynamic approximation, which for the M2 brane theory requires
$B/T^2 \lesssim 10$ \cite{HartnollHerzog}. Outside this regime one
does not expect (\ref{eq:nernsthydro}) to hold. Although the M2
brane theory can be studied beyond the hydrodynamic regime
\cite{HartnollHerzog}, in that case our arguments in section 3.2 for combining
the effect of impurities and the magnetic field do not hold.

The qualitative form of figure \ref{fig:contournorho}a is
determined from the fact that $N \sim 1/F(B/T^2)$, with $F$ a
monotonically decreasing function. This scaling will be a
universal feature of the Nernst effect at $\rho = 0$ in the
vicinity of a quantum critical point. The scaling near a general
quantum critical point will be
\be\label{eq:Nscaling}
N \sim \frac{T^{2 - 2\Delta_\ocal}}{F(B/T^2)} \,.
\ee
We illustrate this general form in figure
\ref{fig:contournorho}b, taking $F(x) = 1/(1+x^2)$ for concreteness.
It is interesting to note that this
qualitative form appears to arise in Fig.~3b of
ref.~\cite{ardavan} at low magnetic fields and just above the
superconducting phase transition temperature, $T_c$. The authors
of ref.~\cite{ardavan} study organic superconductors, and their
Fig.~3b is, like our figure
\ref{fig:contournorho}, a plot of $N(T,B)$. Their system also has
zero charge density, as it is not doped. The region of interest
(low $B$, just above $T_c$) is precisely that for which
\cite{ardavan} propose that strongly correlated electron physics
is important, as are the effects of a nearby Mott transition.

We can try to compute the Nernst signal with a finite charge
density using the formula (\ref{eq:nernsthydro}). As a benchmark,
one can compare with experimental results for the underdoped
cuprate superconductors. It was argued in \cite{Hartnoll:2007ih}
that the charge density was given by the difference in the doping
from the commensurate insulating state at $x_I = 1/8$, divided by
the area of a unit lattice cell. For underdoped LSCO, at $x - x_I
= -0.025$ for example, the lattice constant is $a = 3.78\, \ang$,
leading to
\be\label{eq:rho0}
\rho = \frac{e (x-x_I)}{a^2} \approx 0.028 \frac{\text{C}}{\text{m}^2}
\,.
\ee
The conductivity at the critical point can be estimated to be
\cite{liu}
\be\label{eq:sigma0}
\sigma_0 \approx \frac{4 e^2}{h} \approx 1.6 \times 10^{-4} \frac{\text{C}^2}{\text{J
s}}\,,
\ee
allowing us to obtain
\be\label{eq:rho}
\frac{\rho}{\sigma_0} \approx 175 \, \text{tesla} \,.
\ee
We have expressed the result in tesla, so that it may be compared
to the value of the background magnetic field. The value of 175
tesla is significantly larger than the typical magnetic fields
applied in experiments measuring the Nernst effect in the cuprate
superconductors. In our expression for the Nernst coefficient, the
magnetic field $B$ only appears in the combination $B^2 +
\rho^2/\sigma_0^2$. Therefore the large value of the charge
density in (\ref{eq:rho}) swamps out the $B$ dependence, and the
resulting plot of $N(T,B)$ shows simply vertical lines. We might
note however that both (\ref{eq:rho0}) and (\ref{eq:sigma0}) are
only tentative identifications, so the result (\ref{eq:rho}) could
change substantially. In particular, the conductivity
(\ref{eq:sigma0}) assumes the putative nearby
superconductor-insulator quantum phase transition in the cuprates
is in the same universality class as the films reviewed in
\cite{liu}.

It was suggested in \cite{Hartnoll:2007ih} that the quantity that
should be compared with measurements of the Nernst effect in the
cuprates is not $N$, but rather the off diagonal thermoelectric
coefficient $\alpha_{xy}$. This was to isolate the contribution of
critical superconducting fluctuations from non-critical fermionic
contributions to the condutivity. This coefficient gives the
electrical current generated by an applied thermal gradient: $J_x
= - \alpha_{xy}
\pa_y T$. It is related to the Nernst response by the electrical
conductivity, roughly $\alpha_{xy} \approx
\sigma_{xx} B N$. Here $\sigma_{xx}$ is the full conductivity of the system,
not just that of the critical fluctuations. The thermoelectric
coefficient has a $B$ dependence that is not swamped out by a
large charge density
\cite{Hartnoll:2007ih}, and indeed reproduces various
features of the experimental data for $N(T,B)$ in the cuprates.
The large charge density implies that if we use our M2 brane
expressions for the impurity timescale
$\tau_\text{imp}(\rho,B,T)$, we will not introduce any extra
dependence on $B$ into the previous results of
\cite{Hartnoll:2007ih}. In any case, the high critical temperature
of the cuprates combined with the value for the velocity $v$,
discussed above, implies that $B/T^2,
\rho/T^2 \ll 1$ over the region of interest. Therefore it is
consistent to neglect the $B$ and $\rho$ dependence of
$\tau_\text{imp}$ and the $\epsilon + P$ for these systems.

\begin{figure}[h]
\begin{center}
(a) \epsfig{file=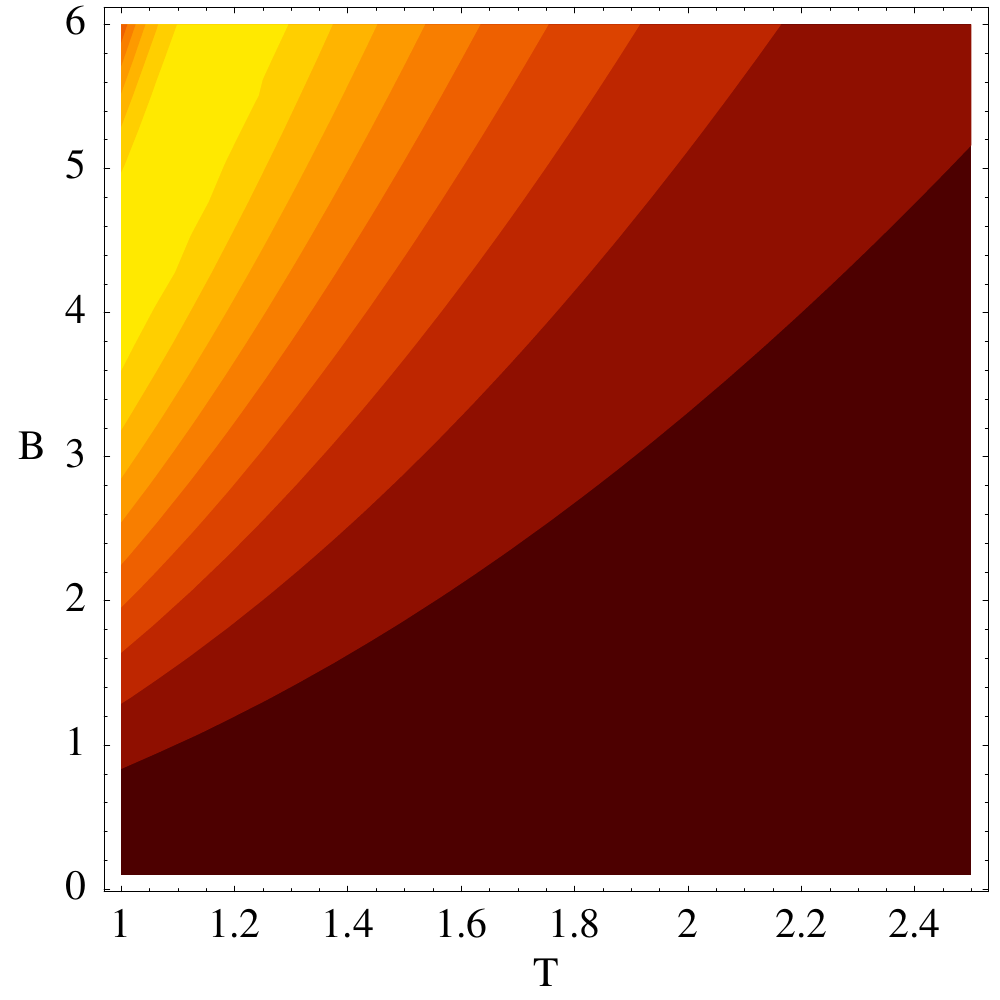,width=6cm}%
(b) \epsfig{file=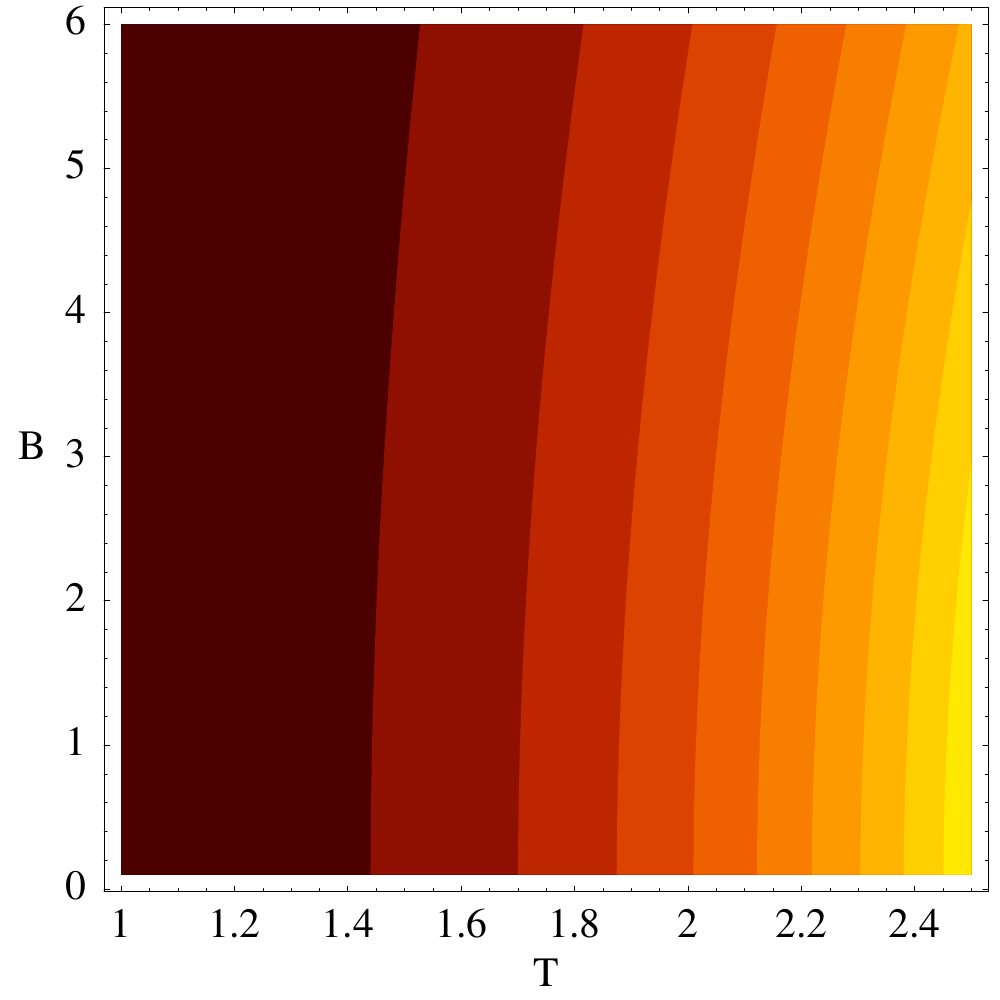,width=6cm}
\end{center}
\caption{ The Nernst signal $N$ for the truncated M2 brane
theory as a function of $B$ and $T$, with (a) charge density
$\rho/\sigma_0 = 8.3 \times 10^{-5}$ and (b) charge density $\rho/\sigma_0 =
10$, and $\bar V = 0.1$ in both cases. The impurity potential is
coupled to $\ocal_+$. Lighter denotes a larger Nernst
coefficient.}
\label{fig:contourrhosmall}
\end{figure}

In figure \ref{fig:contourrhosmall} we have plotted the Nernst
coefficient for the truncated M2 brane theory with $\rho/\sigma_0
= 8.3 \times 10^{-5}$ and $\rho/\sigma_0 = 10$, in each case with
$\bar V = 0.1$. We have again used $\ocal_+$ as the operator
coupling to the impurity potential. The plots are significantly
different and less universal than the vanishing charge density
case, of figure \ref{fig:contournorho}. As $\rho$ becomes nonzero,
a local maximum appears at small temperatures and moves rapidly to
the right as the charge density is increased. Note that unlike the
$\rho = 0$ case we considered previously, plot
\ref{fig:contourrhosmall}a in particular is vulnerable to the
unknown $\bar V^2 B^2$ corrections to $\gamma$ that we mentioned
in section \ref{sec:mag} above. Its qualitative form however
should be correct, as it needs to interpolate between figures
\ref{fig:contournorho} and \ref{fig:contourrhosmall}b.

What we learn from these plots is that (with $\bar V = 0.1$) a
very small $\rho/\sigma_0 \sim 10^{-5}$ is already sufficient to
produce significant deviations from the $\rho = 0$ result. This
suggests that one would have to tune very close to the insulating
doping of $x_I = 1/8$ in order to see the scaling $\rho = 0$
behaviour of equation (\ref{eq:Nscaling}). However, it is curious
that the observations in \cite{ong2} do appear to show similar
behaviour at low $B$ and with $T$ just above $T_c$.

\section{Discussion}

\subsection{Summary of results for the impure CFT}

In this paper we have derived a general formula (\ref{eq:tauimp})
for the impurity relaxation timescale at a quantum critical point.
Some of our steps in section 3 were adapted from earlier works
\cite{gotze, giamarchi, argyres}. We then used our formula to
obtain $\tau_\text{imp}$ for the truncated M2 brane theory. We
found that there were two relevant operators $\ocal_\pm$ that were
neutral under a global $U(1)$ symmetry. This led to the following
results

\begin{itemize}

\item For the `magnetic' operator dual to the scalar field $\psi_+$, increasing
the magnetic field $B$ and charge density $\rho$ suppresses
momentum relaxation due to impurities, see figure
\ref{fig:norho}a. At small $B$ and sufficiently large $\rho$, this
implies that the cyclotron resonance described in
\cite{Hartnoll:2007ih, HartnollHerzog} may be easier to observe
experimentally than previously estimated because the impurity
damping effects become smaller.

\item For the operator dual to the scalar field $\psi_-$, in contrast,
the inverse relaxation time increases when the charge density
$\rho$ and magnetic field $B$ are increased. Beyond the critical
value ${\mathcal Q} = \sqrt{B^2 + \rho^2/ \sigma_0^2} \approx 21
T^2$, the inverse relaxation time actually diverges. See figure
\ref{fig:norho}b. We saw that the divergence is associated with a
new instability in the underlying finite temperature CFT. The
scalar field $\psi_-$ develops an unstable mode when the
${\mathcal Q}$ gets too large. This instability is dual to a phase
transition to an ordered state in which $\ocal_-$ condenses.

\item The Nernst coefficient for the truncated M2 brane theory with vanishing
charge density, $\rho = 0$, is shown in figure
\ref{fig:contournorho} with impurity relaxation due to the
$\ocal_+$ operator. We noted that our dependence on $T$ and $B$ is
qualitatively similar to the results for low magnetic fields and
for temperature close to the critical temperature observed in
various types of superconductors \cite{ardavan, ong2, behnia2}.
This comparison makes most sense for the data in \cite{ardavan},
in which the system is not doped. It is interesting to note,
however, the similarity in the Nernst response between the
different superconductors in this regime.

\end{itemize}

\subsection{Some directions for future work}

Our results are perturbative in the strength of the impurity
potential $\bar V$ and valid at high temperatures. We
furthermore used the memory function formalism, which requires
$1/\tau_\text{imp}$ not to be too large. Various techniques have
been developed in condensed matter theory for treating random
disorder more exactly. It would be very interesting to see if
these can be applied to the M2 brane theory.

Our treatment of the M2 brane theory was not completely satisfactory.
We made an inconsistent truncation in order to avoid dealing with
more than one $U(1)$ gauge field.  We also made a nonconventional
choice for the conformal dimension of our supergravity mode $\chi$.
One clear project for the future is to repeat the calculations
presented here including the extra $U(1)$ gauge field and using
the standard choice of conformal dimension for $\chi$.

But it would also be interesting to take a more general
approach.  While
in our M2 brane theory, the operators we considered had conformal
dimension $\Delta_\ocal = 1$, in general the spectrum of operator
dimensions will depend on the CFT. Our bulk equation of motion
(\ref{eq:phieqn}) for the scalar field $\phi$ dual to the operator
$\ocal$ can be generalised to allow the scalar field to have an
arbitrary mass. This is dual to allowing $\Delta_\ocal$ to vary.
One could imagine using the AdS/CFT correspondence more
`phenomenologically' to see how the effects of impurity relaxation
depend on the dimension of $\ocal$.

One result of this paper is the instability of the truncated M2
brane theory at $\sqrt{B^2 +
\rho^2/ \sigma_0^2} \approx 21 T^2$. We noted that this
instability is a counterexample to the original formulation of the
correlated stability conjecture, as the black hole is
thermodynamically stable at this value of the charge density and
magnetic field. As with the previously studied counterexamples
\cite{FGM}, the mismatch between dynamics and thermodynamics
occurs because there is no conserved charge associated with the
scalar fields. The instability presumably indicates the existence
of new dyonic black hole solutions in which both the scalar $\phi$
and pseudoscalar $\chi$ are nonvanishing. 
It would be
interesting to find these new solutions and understand their
implications for the dual theory.

In combining the effects of impurities and an external magnetic
field, we were constrained to work in a regime where $B/T^2$ was
not too large. It would be nice to relax this requirement. An
approach via Ward identities looks promising.

We have noted that the Nernst effect in systems with vanishing
charge density can be a very clean indicator of the proximity of a
quantum critical point, via the scaling relation
(\ref{eq:Nscaling}). Perhaps this scaling can be searched for in
future measurements of the Nernst effect.

\section*{Acknowledgements}

We would like to thank Ofer Aharony, David Berenstein, Steve
Gubser, David Huse, Igor Klebanov, Nai-Phuan Ong, Joe Polchinski
and Antonello Scardicchio for valuable discussions. We would also
like to thank Subir Sachdev and Markus Mueller for helpful
correspondence. SAH thanks the Weizmann Institute for hospitality
while this work was being completed. This work was supported in
part by the National Science Foundation (NSF) under Grants No.\
PHY-0243680 and PHY05-51164. Any opinions, findings, and
conclusions or recommendations expressed in this material are
those of the authors and do not necessarily reflect the views of
the NSF.

\appendix

\section{Relating $G^R_{\pcal\pcal}$ to a conductivity}
\label{sec:transport}

The connection between $G^R_{\pcal \pcal}(\omega, 0)$ and $\theta$
is established as follows. The momentum conductivity is defined by
\be\label{eq:thetadef}
\langle \pcal \rangle = - \theta(\w,0) \nabla T \,.
\ee
The momentum density $\pcal$ can be sourced by fluctuating the
background spacetime metric. Specifically, by definition there is
a coupling between metric fluctuations and the stress tensor in
the action, $\delta S = \frac{1}{2} \int d^3x  \, T^{\mu\nu}
\delta g_{\mu\nu}$. (We are treating $T^{\mu\nu}$ as a tensor
density rather than a tensor field and hence have absorbed a
factor of $\sqrt{-g}$.) Without loss of generality by rotation
invariance, we can restrict to a fluctuation only in the spatial
direction $\delta g_{0i}$. Choosing $n^j = \delta_{ij}$, linear
response theory then implies
\be\label{eq:thetaGR}
\langle \pcal \rangle =
G^R_{\pcal \pcal}(\omega, 0) \delta g_{0 j} n^j \,.
\ee
The remaining step is to show that $\delta g_{0i}$ is gauge
equivalent to a thermal gradient. Recall that under a
diffeomorphism generated by the vector $\xi^a$, metric
perturbations transform as $\delta g_{a b} = \pa_a \xi_b + \pa_b
\xi_a$. We choose $\xi_i = 0$ and $\pa_i \xi_0 = - \delta g_{0
i}$. After this gauge transformation $\delta g_{0 i}$ vanishes and
$\pa_i \delta g_{00} = 2 i \w \delta g_{0 i}$, assuming the
fluctuations have a time dependence of the form $e^{-i\omega t}$.
Recall that the Euclidean time direction is periodic with period
$1/T$. It is convenient to fix the period and take $g_{00} =
1/T^2$. A shift in temperature thus implies $\pa_i g_{00} = -
2\pa_i T/T^3 = - 2g_{00}
\pa_i T /T$, to leading order in $\pa_i T$.
Putting these formulae together leads to the intermediate result
\be\label{eq:interg0i}
\delta g_{0i} = - \frac{g_{00} \pa_i T }{i \omega T} \ .
\ee
We have to be careful in interpreting this result because we
rescaled time at an intermediate step.  Moreover, we rescaled
$g_{00}$ without rescaling $\omega$. We now need to rescale time
back to the lab frame, where time has period $1/T$ in the
Euclidean direction, being careful not to rescale $\omega$:
\be
\delta g_{0i} = - \frac{\pa_i T}{i \omega T} \ .
\ee
Thus we find from (\ref{eq:thetadef}) and (\ref{eq:thetaGR}) that
\be
i \theta(\omega,0) T \omega \equiv G^R_{\pcal \pcal}(\omega, 0)
\,,
\ee
as claimed in the main text. As we noted, $G^R_{\pcal
\pcal}(\omega, 0)$ needs to be corrected by a contact term to give
a sensible $\w \to 0$ limit for the conductivity.

\section{Hydrodynamics and the memory method}
\label{sec:hydrotau}

In this appendix we show that the results obtained from
hydrodynamics in \cite{HartnollHerzog} and \cite{Hartnoll:2007ih}
are completely consistent with the memory function methods.
Consider $G_{\jcal\jcal}^R \equiv \omega \sigma_+$ for the choice
$n_i = (1,-i)/\sqrt{2}$. We will henceforth suppress the
superscript $R$ since it is clear we work only with retarded
Green's functions in this appendix. In
\cite{HartnollHerzog}, for a translationally invariant theory in
the hydrodynamic limit, it was argued that this holomorphic
conductivity takes the form
\be
\sigma_+ = i \sigma \frac{\omega + i \omega_c^2/\gamma + \omega_c}{\omega + i \gamma - \omega_c} \ .
\ee
Using (\ref{wardjj}), we find that
\be
G_{\pcal \pcal}(\omega) =
(\epsilon + P) \frac{ i\gamma - \omega_c} {\omega + i \gamma - \omega_c} \ .
\ee
As expected, $G_{\pcal \pcal}(0) = \epsilon+P = \chi_0$.
Moreover, this result is entirely consistent with the memory function formalism,
\be
\lim_{B \to 0} \frac{\omega G_{PP}(\omega)}{\chi_0} = i \gamma - \omega_c \ .
\ee
This limit is $B/T^2 \to 0$. The frequency $\w/T$ is kept small
but fixed.

In \cite{Hartnoll:2007ih}, it was shown that an impurity
scattering time could be introduced by making the replacement
$\omega \to \omega + i/\tau$ in the conductivities. This followed
from hydrodynamics plus a relaxation of momentum imposed by hand.
This substitution yields
\be
\sigma_+' = i \sigma \frac{\omega + i/\tau + i \omega_c^2/\gamma + \omega_c}
{\omega + i/\tau + i \gamma - \omega_c} \ .
\ee
which agrees with the combination $\sigma_{xy} + i \sigma_{xx}$ from
\cite{Hartnoll:2007ih}.  We use the superscript $'$ to indicate a Green's function or
conductivity in the presence of impurities.
The prescription is trickier when applied to $G_{\pcal \pcal}$.  The relevant transport coefficient
we will call $\theta_+$ is related via $\omega T \theta_+ = G_{\pcal \pcal}(\omega) - \chi_0$.  Thus
\be
\theta_+' T = - \frac{\chi_0}{\omega + i/\tau + i\gamma - \omega_c} \ ,
\ee
which then implies that
\be
G_{\pcal \pcal}'(\omega) = \chi_0 \frac{i/\tau + i\gamma - \omega_c}{\omega + i/\tau + i\gamma - \omega_c} \ .
\ee
Again, the memory function method works with this modified $G_{\pcal \pcal}'$ yielding
a pole at the appropriate $\omega = - i/\tau - i \gamma + \omega_c$.

Let's now compute
\be
B^2(\sigma_+'(\omega) - \sigma_+(0) )= \chi_0\frac{ (\omega + i/\tau)(i \gamma-\omega_c)}
{(\omega + i/\tau + i \gamma - \omega_c)} \ ,
\label{hmmph}
\ee
and check the extent to which (\ref{wardjj}) continues to hold in the small $1/\tau$ and $B$ limit:
\be
\lim_{B,\frac{1}{\tau} \to 0} \frac{\omega}{\chi_0} \left(G_{\pcal \pcal}'(\omega) - \frac{B^2}{\omega^2}(G_{\jcal \jcal}'(\omega)-G_{\jcal\jcal}(0))\right)
=
i \gamma - \omega_c  +i/\tau - (i\gamma - \omega_c) = i/\tau \ .
\ee
Indeed, that's exactly the result we were looking for, suggesting
(\ref{eq:total}) is correct. 
We have made only an indirect assumption about
the $B$ dependence of $1/\tau$.  Our hydrodynamic
framework is valid only up to $\ocal(B^2)$.  
Thus while (\ref{eq:timescale}) appears to capture
the full $B$ dependence of $1/\tau$, in this 
hydrodynamic context, we can trust the answer only
up to $\ocal(B^2)$.
%

\section{Action for fluctuations without truncation}
\label{sec:G}

In this appendix we give, for possible future use, the untruncated
action for perturbations of the dyonic black hole background
(\ref{eq:metric}) and (\ref{eq:Ffield}).

From appendix F of \cite{Duff} one can check that the full action
for 3 scalars, 3 pseudoscalars and 4 gauge fields may be
consistently truncated for linearised perturbations of the dyonic
black hole to one scalar, $\phi$, one pseudoscalar, $\chi$, and
one gauge field $G$:
\bea\label{eq:actionfull}
S  & = \displaystyle -\frac{1}{2\kappa_4^2} \int d^4 x \sqrt{-g} &
\Big( R -
\half \left[ (\partial_\mu \phi) (\partial^\mu \phi) + (\partial_\mu \chi) (\partial^\mu \chi) \right]
+ 2 L^{-2} \left[3 +
\half (\phi^2 + \chi^2)  \right] \nonumber \\
 & &
 - L^2 \left[1 + \half (\phi^2 - \chi^2)  \right] F_{\mu\nu} F^{\mu\nu} + \half L^2 \phi \chi
 \epsilon^{\mu \nu \rho \sigma} F_{\mu\nu} F_{\rho\sigma}  \nonumber \\
 & &
 - L^2 G_{\mu\nu} G^{\mu\nu}
- 2 L^2 \phi F_{\mu\nu} G^{\mu\nu}  +  L^2 \chi
\epsilon^{\mu\nu\lambda \rho} F_{\mu\nu} G_{\lambda \rho}
 \Big) \, .
\eea
We have used the same notation as in the main text. In particular
$G_{\mu \nu} = \delta F_{\mu \nu}^{(1)} + \delta F_{\mu
\nu}^{(2)} - \delta F_{\mu \nu}^{(3)} - \delta F_{\mu \nu}^{(4)}$,
whereas the background gauge field is $F_{\mu \nu} = F_{\mu
\nu}^{(1)} + F_{\mu \nu}^{(2)} + F_{\mu \nu}^{(3)} + F_{\mu \nu}^{(4)}$.

Once again it is useful to introduce the fields $\psi_\pm$ defined
in (\ref{eq:psi}). Evaluated on the dyonic black hole background,
the action becomes
\bea
S &  = \displaystyle - \frac{1}{2 \kappa_4^2} \int d^4x \sqrt{-g}
&\Big(
-\half [ (\partial_\mu \psi_+)(\partial^\mu \psi_+) +
(\partial_\mu \psi_-) (\partial^\mu \psi_-)]
+ L^{-2} (\psi_+^2 + \psi_-^2)\\
&&
- L^{-2}  z^4 (h^2+q^2)  (\psi_+^2-\psi_-^2) - L^2 G_{\mu\nu} G^{\mu\nu}  \\
&& -4 L^{-2} \frac{z^4}{\alpha^2} \sqrt{h^2+q^2}
\left(  \alpha \psi_-  G_{tz}  + \psi_+ G_{xy} \right)   \Big) \ .
\eea
In this expression we note that once again the background magnetic
and electric charges only appear in the combination $h^2+q^2$. It
should be possible to decouple the equations of motion following
from this action and find the momentum relaxation in the full
theory. Solving the full equations would also be of interest in
terms of revisiting the computations in \cite{Gubser:2000mm,
Gubser:2000ec}.

\section{Exponential falloff at large $k$}
\label{sec:WKB}

The exponential falloff of the imaginary part of the Green's
function with large spatial momentum $k$ can be seen from a WKB
analysis. To perform a WKB analysis of the differential equation
(\ref {eq:phieqn}), we first transform it into Schrodinger form.
(A very similar analysis was performed in an appendix of
ref.~\cite{Son:2002sd}.) To that end, we define a new wavefunction
$\Phi_\pm$ such that $\Phi_\pm (\rho) = \psi_\pm(z) /z$ and a new
radial coordinate $\rho(z)$ such that $\partial_z \rho = 1/f$.
With these new definitions, the differential equation becomes
\be
(-\partial_\rho^2 + V(\rho) - \fw^2) \Phi_\pm (\rho) = 0
\ee
where
\be
V(\rho) = \frac{f}{z^2} (m^2 L^2 + \fk^2 z^2 +
2 f - z \partial_z f
\pm 2 z^4 (q^2 + h^2)
) \ .
\ee
In terms of the new radial variable, we can take
\be
\rho = \int_0^z \frac{dz'}{f(z')} \ .
\ee
Thus, the horizon at $z_h=1$ has been moved to $\rho_h \to
\infty$. The boundary at $z_b=0$ remains at $ \rho_b=0$.

Note that for $m^2L^2=-2$ we have $V(0) = \fk^2$. We are
interested in the case of spacelike momentum, $k \gg \omega$, for
which there are no classical turning points. The wavefunction is
exponentially damped everywhere. The imaginary part of the
retarded Green's function will be proportional to the probability
for the particle to reach the horizon at $z_h=1$:
\be
\mbox{Im} \, G^R_{\ocal \ocal} \sim \exp \left(-2 \int_0^ \infty d\rho \, \sqrt{V(\rho) - \fw^2} \right) \ .
\ee

\begin{figure}[h]
\begin{center}
\epsfig{file=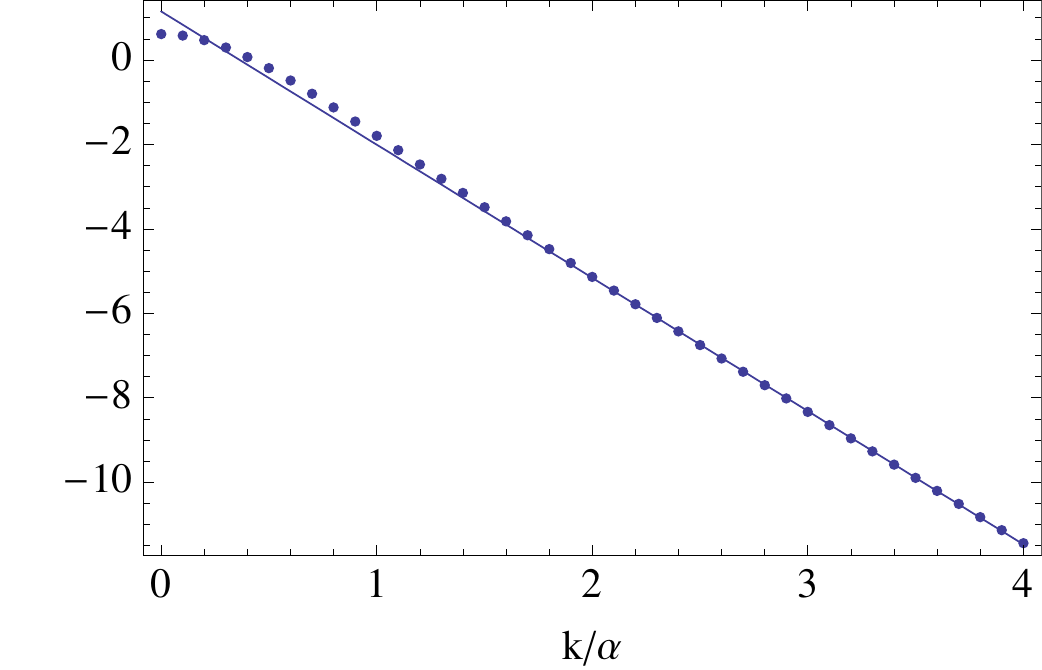,width=7cm}%
\end{center}
\caption{ The points are numerically determined values
of the log of the imaginary part of $G^R_{\ocal \ocal}(0,k)$ as a
function of $k$ for $q^2+h^2=1/4$ and the $\psi_+$ scalar field.
More specifically, we plot
$\ln (\lim_{\fw \to 0} B/ \fw A )$. The line is a best fit $1.15 - 3.15 \fk$.
For these values of $q$ and $h$, WKB gives us $a=2.86$.}
\label{fig:expdamping}
\end{figure}

Given that $k \gg \omega$, we can approximately evaluate this
integral to give $\mbox{Im} \, G^R_{\ocal \ocal} \sim \exp(-a k /
\alpha)$ where we have defined the coefficient
\be
a \equiv 2 \int_0^1 \frac{dz}{ \sqrt{f(z)}} \ .
\ee
In other words, the WKB analysis is telling us that $\mbox{Im}\,
G^R_ {\ocal \ocal}$ is exponentially damped for large enough
$k/\alpha$. In the case where $h=q=0$, we may evaluate the integral
analytically to find
\be
\int_0^1 \frac{dz} {\sqrt{1-z^3}} = \frac{\sqrt{\pi} \Gamma(4/3)} {\Gamma(5/6)} \approx 1.402 \ .
\ee
In order for our integral (\ref{eq:F1}) over $\fk=k/\alpha$ of the
imaginary part of the Green's function to converge, this
exponential damping is important. We observed it numerically for
more general values of $h$ and $q$. In Figure
\ref{fig:expdamping}, we plot the log of the imaginary part of
$G^R_{\ocal \ocal}(0,k)$ as a function of $k/\alpha$ and indeed see linear
behavior at large $k$.

\end{document}